%% file: Maestro_main.tex
\documentclass[lettersize,journal]{IEEEtran}
\usepackage[LGR,T1]{fontenc}
\usepackage{amsmath,amsfonts}
\usepackage{algorithmic}
\usepackage{algorithm}
\usepackage{array}
\usepackage{url}
\usepackage{verbatim}
\usepackage{graphicx}
\usepackage{cite}
\usepackage{svg}
\usepackage{orcidlink}
\usepackage{xcolor}
\usepackage{acronym}
\usepackage[per-mode=symbol]{siunitx}
\usepackage{wrapfig} 
\usepackage{booktabs}
\usepackage{multirow}
\usepackage{threeparttable} 
\usepackage{paralist}

\usepackage{tabularx}
\usepackage{rotating}

\DeclareSIUnit\flops{FLOPS}
\DeclareSIUnit\gflops{GFLOPS}
\DeclareSIUnit\gflopsw{GFLOPS/W}
\DeclareSIUnit\gops{GOPS}
\DeclareSIUnit\gopsw{GOPS/W}
\DeclareSIUnit\Mbps{Mb/s}

\DeclareSIUnit\flop{FLOP}
\DeclareSIUnit\cycle{cycle}
\DeclareSIUnit\cycles{cycles}
\DeclareSIUnit\GE{GE}
\DeclareSIUnit\kb{kB}
\DeclareSIUnit\kb{KiB}

\acrodef{BLE}{Bluetooth Low Energy}
\acrodef{HW}{Hardware}
\acrodef{PC}{Personal Computer}
\acrodef{MCU}{Microcontroller Unit}
\acrodef{ASIC}{Application-Specific Integrated Circuit}
\acrodef{IoT}{Internet-of-Things}
\acrodef{AI}{Artificial Intelligence}
\acrodef{SIMD}{Single-Instruction-Multiple-Data}
\acrodef{SCM}{Scratchpad Memory}
\acrodef{ISA}{Instruction-Set Architecture}
\acrodef{HPC}{High-Performance Computing}
\acrodef{HMI} {Human-Machine Interface}
\acrodef{HMIs} {Human-Machine Interfaces}
\acrodef{HWPE}{Hardware Processing Engine}
\acrodef{ML}{Machine Learning}
\acrodef{FIR}{Finite Impulse Response}
\acrodef{IIR}{Infinite Impulse Response}
\acrodef{FFT}{Fast Fourier Transform}
\acrodef{DFT}{Discrete Fourier Transform}
\acrodef{iFFT}{Inverse Fast Fourier Transform}
\acrodef{SVM}{Support Vector Machines}
\acrodef{KNN}{K-Nearest Neighbors}
\acrodef{LDA}{Linear Discriminant Analysis}
\acrodef{DSP}{Digital Signal Processing}
\acrodef{SoC}{System-on-a-Chip}
\acrodef{sEMG}{surface ElectromMyoGraphy}
\acrodef{FPU}{Floating Point Unit}
\acrodef{FPUs}{Floating Point Units}
\acrodef{IC}{Integrated Circuit}
\acrodef{FMA}{Fused Multiply-Accumulate}
\acrodef{FMAs}{Fused Multiply-Accumulate}
\acrodef{APB}{Advanced Peripheral Bus}
\acrodef{AXI}{Advanced eXtensible Interface}
\acrodef{VTU}{Vector-Tensor Unit}
\acrodef{VU}{Vector Unit}
\acrodef{TU}{Tensor Unit}
\acrodef{FLLs}{Frequency-Locked Loops}
\acrodef{FLL}{Frequency-Locked Loop}
\acrodef{TCDM}{Tightly-Coupled Data Memory}
\acrodef{DMA}{Direct Memory Access}
\acrodef{CDC}{Clock Domain Crossing}
\acrodef{FU}{Functional Unit}
\acrodef{FUs}{Functional Units}
\acrodef{VFU}{Vector Functional Unit}
\acrodef{VFUs}{Vector Functional Units}
\acrodef{VLEN}{Vector Length}
\acrodef{VAU}{Vector Arithmetic Unit}
\acrodef{VLSU}{Vector Load Store Unit}
\acrodef{VSLDU}{Vector Slide Unit}
\acrodef{IPU}{Integer Processing Unit}
\acrodef{VRF}{Vector Register File}
\acrodef{GEMM}{General Matrix-Matrix Multiplication}
\acrodef{GEMM-Ops}{General Matrix-Matrix Operations}
\acrodef{PEs}{Processing Elements}
\acrodef{CEs}{Computing Elements}
\acrodef{CE}{Computing Element}
\acrodef{VFMACC}{Vector Fused Multiply-Accumulate}
\acrodef{TCSR}{Tensor Control Status Register}
\acrodef{CSR}{Control Status Register}
\acrodef{LMUL}{Length Multiplier}
\acrodef{LUTs}{Look-Up Tables}
\acrodef{LUT}{Look-Up Table}
\acrodef{DIT}{Decimation In Time}
\acrodef{SDOTP}{Sum of Dot Product}
\acrodef{US}{Ultrasound}
\acrodef{WUS}{Wearable Ultrasound}
\acrodef{TGC}{Time Gain Compensation}
\acrodef{GF}{Gaussian Filtering}
\acrodef{FC}{Fully Connected}
\acrodef{MP}{Max Pooling}
\acrodef{AVG}{Average Pooling}
\acrodef{CNN}{Convolutional Neural Network}
\acrodef{DNN}{Deep Neural Network}
\acrodef{HT}{Hilbert Transform}
\acrodef{ULP}[ULP]{Ultra-Low Power}
\acrodef{RMSE}[RMSE]{Root Mean Squared Error}
\acrodef{PCA}[PCA]{Principal Component Analysis}
\acrodef{BN}[BN]{Batch Normalization}
\acrodef{RVV}[RVV]{RISC-V Vector}
\acrodef{EEG}[EEG]{Electroencephalography}
\acrodef{EMG}[EMG]{Electromiography}

\newcommand{\mycomment}[1]{}
\begin{document}

\title{Maestro: A 302 GFLOPS/W and 19.8GFLOPS RISC-V Vector-Tensor Architecture for Wearable Ultrasound Edge Computing}

\author{\IEEEauthorblockN{
     Mattia Sinigaglia \orcidlink{0000-0002-3350-8789},
     Amirhossein Kiamarzi \orcidlink{0009-0008-7092-6227},
     Marco Bertuletti \orcidlink{0000-0001-7576-0803},~\IEEEmembership{Student Member,~IEEE},
     Luigi Ghionda \orcidlink{0009-0000-5748-1085},
     Mattia Orlandi \orcidlink{0000-0002-8553-3273}~\IEEEmembership{Student Member,~IEEE},
     Riccardo Tedeschi \orcidlink{0009-0007-4483-9261},
     Aurora Di Giampietro \orcidlink{0009-0000-4821-4364},
     Yvan Tortorella \orcidlink{0000-0001-8248-5731},
     Luca Bertaccini \orcidlink{0000-0002-3011-6368},~\IEEEmembership{Member,~IEEE},
     Simone Benatti \orcidlink{0000-0002-5700-5342}~\IEEEmembership{Member,~IEEE},
     Giuseppe Tagliavini \orcidlink{0000-0002-9221-4633},~\IEEEmembership{Senior Member,~IEEE},
     Luca Benini \orcidlink{0000-0001-8068-3806}, ~\IEEEmembership{Fellow,~IEEE},
     Francesco Conti \orcidlink{0000-0002-7924-933X}, ~\IEEEmembership{Senior Member,~IEEE},
     and Davide Rossi \orcidlink{0000-0002-0651-5393}, ~\IEEEmembership{Senior Member,~IEEE}}
     
     
     \thanks{ Manuscript submitted March 6, 2025.}
     \thanks{
     Mattia Sinigaglia, Amirhossein Kiamarzi, Luigi Ghionda, Mattia Orlandi, Riccardo Tedeschi, Aurora Di Giampietro,  Yvan Tortorella, Giuseppe Tagliavini,  Luca Benini, Francesco Conti, and Davide Rossi are with the Department of  Electrical, Electronic and Information Engineering (DEI), University of Bologna, 40136 Bologna, Italy.
     Simone Benatti is with the Department of Science and Methods for Engineering (DISMI), University of Modena e Reggio Emilia, Italy.
     Marco Bertuletti, Luca Bertaccini and Luca Benini are with the Integrated Systems Laboratory (IIS), ETH Zürich, 8092 Zürich, Switzerland.
     }
     \thanks{
     This work was supported in part through the ISOLDE (101112274) and TRISTAN (101095947) project that received funding from the HORIZON CHIPS-JU program and in part by the Spoke 1 on Future High-Performance-Computing (HPC) of the Italian Research Center on High-Performance Computing, Big Data and Quantum Computing (ICSC) that received funding from the Ministry of University and Research (MUR) for the Mission 4–Next Generation EU program.
     }
}

\markboth{}%
{Shell \MakeLowercase{\textit{et al.}}: A Sample Article Using IEEEtran.cls for IEEE Journals}


\maketitle

\bstctlcite{IEEEexample:BSTcontrol}

\input{00-Abstract.tex}


\input{01-Introduction.tex}

\input{02-Related_Work}

\input{03-Maestro_Architecture}


\input{04-Physical_Implementation_and_measurements}

\input{05-End-to-end_Ultrasound_Application}

\input{06-SoA_Comparison}

\input{07-Conclusion}

\bibliographystyle{IEEEtran}
\bibliography{bibliography.bib}

\newpage

\section{Biography}

\input{Authors/Authors_bio_picture}
\vfill

\end{document}

%% file: 00-Abstract.tex
\begin{abstract}
Most Wearable Ultrasound (WUS) devices lack the computational power to process signals at the edge, instead relying on remote offload, which introduces latency, high power consumption, and privacy concerns. 
We present Maestro, a RISC-V SoC with unified Vector-Tensor Unit (VTU) and memory-coupled Fast Fourier Transform (FFT) accelerators targeting edge processing for wearable ultrasound devices, fabricated using low-cost TSMC 65nm CMOS technology. 
The VTU achieves peak 302GFLOPS/W and 19.8GFLOPS at FP16, while the multi-precision 16/32-bit floating-point FFT accelerator delivers peak 60.6GFLOPS/W and 3.6GFLOPS at FP16, 
We evaluate Maestro on a US-based gesture recognition task, achieving 1.62GFLOPS in signal processing at 26.68GFLOPS/W, and 19.52GFLOPS in Convolutional Neural Network (CNN) workloads at 298.03GFLOPS/W. 
Compared to a state-of-the-art SoC with a similar mission profile, Maestro achieves a 5× speedup while consuming only 12mW, with an energy consumption of 2.5mJ in a wearable US channel preprocessing and ML-based postprocessing pipeline.

\end{abstract}

%% file: 01-Introduction.tex
\section{Introduction} \label{sec:Introduction}

\IEEEPARstart{T}{he} recent advancements in \ac{US}~\cite{10529060} transceiver technology are enabling the development of \ac{WUS} systems~\cite{HU2024107401} for the next-generation \ac{HMIs}, with applications such as movement tracking~\cite{9185023} and gesture recognition~\cite{HU2024107401,10208224}, benefitting from an increase in the miniaturization and energy efficiency of \ac{US} probes: correspondingly, a fundamental requirement for all wearable systems is that they must be unobtrusive, compact and cost effective.
Next-generation \ac{WUS} systems also need to reduce the physical footprint of batteries that support onboard processing and ensure low operating temperature without active cooling.
This translates in very strict $<$50mW power constraints, which in turn require an extremely energy-efficient edge processing platform.

For wearable processing systems focusing on biopotential signals (e.g., \ac{EEG}, \ac{EMG}), several energy-efficient SoCs have been proposed for~\cite{ManuelWearable,BioWAP,VEGA}. 
For this class of signals, input bandwidth is in the low-KHz range, preprocessing and conditioning effort is relatively low.  
On the other hand, \ac{US} signals have orders of magnitude larger input bandwidth, and require more complex conditioning to convert A-mode data---i.e., the echo signals from muscle deformations---into meaningful features for downstream tasks (e.g., gesture classification)~\cite{8662605,9872106,10557691,UltrasoundPrediction,MOHIT2024}.
This is partly done by analog  circuitry that excites the transducers, acquires the echoes, and amplifies the signal; and in part in the digital domain, where signals are usually filtered with classical \ac{DSP} algorithms such as \ac{FIR} and \ac{IIR} filters, \ac{HT}, and \ac{FFT} before being fed (often in the frequency domain) to into one of a large variety of \ac{ML} models~\cite{10557691} (both conventional~\cite{9872106}---e.g., \ac{SVM}, \ac{LDA}---and based on deep learning~\cite{10208224}).
The large diversity of workloads involved in WUS processing calls for dedicated architectural solutions that couple flexibility with energy efficiency.

State-of-the-Art \ac{ULP} \acp{MCU}~\cite{ManuelWearable, 7864441,STM32L0}, commonly used in wearable scenarios, prioritize minimal power consumption with respect to performance and are unsuited to the complexity of \ac{WUS} workloads.
On the other hand, modern \ac{AI}-accelerated \acp{MCU} provide advanced \ac{AI} capabilities but generally lack high-performance \ac{DSP} features, particularly for frequency-domain processing~\cite{NDP250, STM32N6,BioWAP, AlifBalletto}.
Whenever a given workload is not directly accelerated, these systems rely on software execution on simple, low-performance RISC processors, which severely hinders end-to-end performance in the case of \ac{WUS}.
Embedded vector processors~\cite{Dabbelt2016VectorPF} provide an attractive alternative to couple performance and efficiency on a wide variety of \ac{DSP} tasks with the flexibility of a programmable architecture.
However, when targeting workloads that include modern \ac{AI} algorithms such as \acp{CNN} or Transformers, vector processors may still require an additional performance boost from an external tensor unit to achieve the target performance, with significant area and power overhead.

In this work, we propose to go one step further in integration, with a \ac{WUS} processing architecture based on a unified vector/tensor unit to boost efficiency on \ac{AI} and linear-algebra-centric workloads.
By merging the tensor unit into the vector processor, we are able to share part of their microarchitectural resources, minimizing the overhead of the tensor unit while still retaining its performance.
Additionally, to enable fast and efficient frequency-domain processing in the \ac{WUS} scenario, we couple our processor with a companion multi-precision floating-point \ac{FFT} engine.
As an embodiment of this architectural approach, we present \textit{Maestro}, a unified Vector-Tensor \ac{WUS} \ac{SoC} prototyped in 65nm CMOS technology.
In detail, the key contributions of this work are the following:
\begin{itemize}
    \item a multi-precision unified \ac{VU}, based on the Spatz architecture~\cite{Cavalcante2023SpatzCC}, supporting the \ac{RVV} \ac{ISA} with FP16--FP64 data, and high-performance \ac{GEMM} operations in FP8--FP16 precision in a fully integrated \ac{VTU};
    \item a multi-precision FP16--FP32 radix-2 \ac{DIT} \ac{FFT} engine employing novel fused-arithmetic dual-output sum-of-dot-products modules for high-precision, low-energy operation;
    \item a heterogeneous compute cluster including the \ac{VU} and \ac{FFT} engine in a memory-coupled configuration with \SI{128}{\kibi\byte} of low-latency L1 memory;
    \item the implementation and characterization of the proposed architecture on a commercial 65nm CMOS technology;
    \item the comparative analysis of the proposed architecture on an end-to-end ultrasound-based gesture recognition application versus GAP9~\cite{VEGA}, a high performance \& energy-efficiency commercial \ac{SoC}.
\end{itemize}
Maestro delivers 1.62GFLOPS and 19.52GFLOPS in frequency domain and \ac{CNN} tasks, respectively, outperforming GAP9 by 2.63$\times$ and 6.53$\times$. It delivers its highest energy efficiency in \ac{CNN} processing at 298GFLOPS/W, 2.19$\times$ higher than GAP9.
Overall, our results show that Maestro's unified Vector/Tensor processor + FFT companion architecture is highly competitive with GAP9 on \ac{WUS} applications, achieving 5$\times$ better end-to-end latency on our target \ac{WUS} workload despite GAP9's much more aggressive technology node (22nm). Furthermore, when employed in a Maestro-based ultrasound system, which acquires and processes data at the edge, the entire system operates with an ultra-low power consumption of just 12mW, achieving an energy cost of only 2.5mJ and operating up to 94 hours.

The manuscript is structured as follows: Section \ref{sec:Related Work} reviews vector processors, hardware accelerators for \ac{AI} and \ac{FFT}, and state-of-the-art \ac{WUS} applications and systems. Sections \ref{sec:Maestro Architecture}, \ref{sec:Unified Vector Tensor Unit}, and \ref{sec:Mixed precision FFT Accelerator} detail Maestro's architecture and accelerators. Section \ref{sec:Physical Implementation} presents the silicon prototype implementation and measurements. Section \ref{sec:End-to-end Ultrasound Application} evaluates the performance, energy efficiency, and latency of an A-Mode \ac{WUS} application. Section \ref{sec:SoA Comparison} compares Maestro with similar prototypes, and Section \ref{sec:Conclusion} concludes with key findings and future research directions.

%% file: 02-Related_Work.tex
\input{02.1-SoA_Table}

\section{Related Work} \label{sec:Related Work}
This section analyzes the design principles and architectural components that define the proposed \ac{SoC}, including vector cores, tensor cores, and \ac{FFT} accelerators. Finally, we review the state of the art in \ac{WUS} applications and systems shown in Table \ref{tab:related-socs}.

\subsection{Vector Cores} \label{subsec:Vector Cores}

Vector processing presents an effective solution to the growing need for programming flexibility coupled with performance and energy efficiency. By leveraging \ac{SIMD} paradigm, a single vector instruction can process multiple data elements, significantly reducing the energy overhead associated with instruction fetch and dispatch~\cite{10.5555/1999263}.

Both ARM and RISC-V instruction sets support advanced vector processing capabilities with dedicated extensions. ARM introduced the Scalable Vector Extension (SVE)~\cite{7924233} and its successor, SVE2, while the RISC-V community developed the RISC-V Vector (RVV) \ac{ISA}~\cite{RVV1}, known for its open and extensible design. The RVV specification reached its frozen 1.0 version in 2021, marking a significant milestone in RISC-V's vector processing capabilities.
Since then, both industry and academia have actively contributed to a wide range of vector processor implementations. These include \ac{HPC} designed for large-scale processing and complex dataset analysis leveraging scalability, efficiency, and parallelism~\cite{Fugaku,ARA,Vitruvius}, \ac{DNN}~\cite{SPEED,9567768,eightCore} and embedded applications~\cite{Dabbelt2016VectorPF, 10631150,Cavalcante2023SpatzCC}.

Ara~\cite{ARA} is the first open-source RVV processor, operating at more than 1GHz in 22FDS Global Foundries technology. It integrates the Linux-capable CVA6 core~\cite{CVA6} with a vector accelerator featuring 16 double-precision lanes, each accessing 16$\times$64-bit elements of the \ac{VRF}. This design delivers 33GFLOPS and 41GFLOPS/W for double-precision matrix multiplication, making it a power-efficient HPC solution with a peak power consumption of 763mW.
Yun~\cite{YUN} is the first silicon prototype based on the smallest Ara configuration, featuring 4$\times$64-bit lanes and a \ac{VLEN} of 4096-bits. It operates up to 280MHz in 65nm TSMC technology and achieves 2.83GFLOPS and 10.8GFLOPS/W for double-precision matrix multiplication, with power consumption ranging from 57mW to 330mW.

Speed~\cite{SPEED}, aims to enable efficient quantized \ac{DNN} inference operating at a frequency of 1.05GHz in TSMC 28nm technology. It supports processing precision ranging from 4-bit to 16-bit and integrates within each of the 4$\times$16-bit lanes an integer-multi-precision (4-8b) tensor unit. Experimental results show that SPEED achieves a peak throughput of 7.37TOPS and an energy efficiency of 1.38TOPS/W for 4-bit operations consuming up to 533mW. 

These architectures are designed for large-scale \ac{HPC} clusters, where their large \ac{VRF} enables efficient compute-intensive edge and cloud systems processing of long vectors across multiple cores. Despite their relatively low per-core power consumption (<1W), their efficiency depends on sustaining high utilization through large-scale parallelism. This makes them well-suited for massively parallel workloads but inefficient for smaller, fragmented tasks, where their computational resources remain underutilized. Moreover, they are designed for application targets with much more relaxed power constraints than \ac{WUS}.

The Spatz architecture ~\cite{Cavalcante2023SpatzCC} operates at over 1GHz in 12LPP technology. It implements an open-source dual-core vector processor based on a cluster that shares 128KB of memory. Each vector core is equipped with a 512-bit \ac{VRF} divided across four 64-bit lanes. This smaller \ac{VRF} design enhances utilization, ensuring more efficient use of functional units, particularly on small workloads, making it well-suited for embedded processing domains. The Spatz cluster achieves a peak performance of 15.7GFLOPS and a peak energy efficiency of 95.7GFLOPS/W when executing a double-precision matrix multiplication workload, consuming up to 164mW.

The cluster proposed in Maestro, built on the Spatz cluster template, extends the vector \ac{DSP} capabilities with frequency domain acceleration by integrating a memory-coupled \ac{FFT} engine and a tightly coupled tensor unit accelerating \ac{ML} and general matrix-matrix operations, offering increased performance and energy efficiency. 

\subsection{Tensor Cores} \label{subsec:Tensor Cores}
\ac{AI} workloads are dominated by linear operations, such as matrix multiplication, making Tensor Core architectures the preferred solution for accelerating these computations. State-of-the-art implementations explore various architectural approaches to optimize performance, power efficiency, and area utilization. We focus the following survey on tensor engines designed for low-power edge applications.

Gemmini~\cite{gemmini} is an open-source systolic array of 256 \ac{CEs} of 8-bit Integer multiply-accumulate designed for inference of \ac{DNN} that support runtime-programmable stationary weight and output data flows.
It has been implemented in Intel 22FFL technology operating at over 961MHz achieving a peak energy efficiency of 73.3GOPS/W. 

Tsunami~\cite{Tsunami}, manufactured in 65nm CMOS technology, targets high energy efficiency in \ac{DNN} training with 1024$\times$16-bit floating-point \ac{CEs} at 200MHz. Relying on pruning to eliminate unnecessary computations, Tsunami delivers 310GFLOPS and 1.71TFLOPS/W on 16-bit floating-point operations with a maximum power consumption of 419mW.

DARKSIDE \cite{Garofalo2023DARKSIDEAH} is a heterogeneous RISC-V compute cluster for TinyML at the extreme edge. It features 8 DSP-enhanced RISC-V cores, a Depth-Wise Convolution Engine, a DataMover, and a 16-bit floating-point tensor core for matrix operations consisting of 32 FP16 \ac{FMA} units. Fabricated in TSMC 65nm, it runs at 290MHz and delivers 18.2GFLOPS at 300GFLOPS/W, with a peak power consumption of 213mW.

Our work, Maestro, introduces a unified vector-tensor processing framework, integrating a tensor unit within the vector core to enable tightly coupled cooperation with the \ac{VRF}. Unlike previous designs~\cite{SPEED,gemmini}, Maestro’s tensor unit, an area-improved design of the open-source RedMulE architecture~\cite{TORTORELLA2023122}, supports mixed-precision 8/16-bit floating-point arithmetic (FP8/FP16) and offers up to 48 \ac{CEs}, enabling efficient embedded processing across diverse workload types by coordinating its execution among other functional units. The inclusion of FP8 and FP16 support balances precision and efficiency for edge \ac{WUS} applications scenarios with \ac{AI} and \ac{DSP} workloads.

\subsection{FFT Accelerators} \label{subsec:FFT Accelerators}
As frequency-domain workloads grow in complexity, software-based \ac{FFT} solutions struggle to achieve real-time performance and energy efficiency, highlighting the need for specialized hardware accelerators.
Wang et al.~\cite{8240277} introduced an \ac{FFT} accelerator capable of handling 24-bit precision fixed-point data, while Sinigaglia et al.~\cite{Echoes} supports 8/16/32-bit fixed-point precision on the TSMC 65nm implementation of a frequency domain Echoes \ac{SoC} delivering up to 200GOPS/W and running up to 350MHz with a power envelope of 134mW.

Wang et al. \cite{8240277} present a 16nm FinFET runtime-reconfigurable $2^n 3^m 5^k$ \ac{FFT} fixed-point accelerator designed for multi-standard wireless communication systems such as LTE and Wi-Fi. The accelerator is built using a memory-based architecture optimized for high area efficiency and is integrated with a RISC-V core to demonstrate a complete software-defined radio (SDR) system operating up to 940MHz, and consuming up to 22.6mW depending on the \ac{FFT} workload. The use of fused floating-point operations tailored for \ac{FFT} processors is explored in studies such as~\cite{5669293,7123672}.

Hitesh et al.~\cite{10509999} investigate the implementation of a high-precision frequency measurement system for MEMS gyroscopes, employing a low-precision (11-bit) floating-point approach for radix-2 \ac{FFT} computation. The system enhances computational efficiency by optimizing memory access patterns through the matrix transposition technique (CTMT), which is applied alongside the Cooley-Tukey \ac{FFT} algorithm, a standard approach for \ac{FFT} computation. The design is capable of performing a 256-point \ac{FFT} while maintaining scalability to variable-length \acp{FFT}. 
Implemented in 65nm CMOS technology, it operates at 100MHz, while delivering a power consumption of approximately 59.1mW at a 1.2V power supply voltage. 

The work by Larry et al.~\cite{9926333} introduces a hardware accelerator specifically designed to enhance the performance of \ac{FFT}W software library for scientific computing.
Fabricated in TSMC 28nm technology, the accelerator adopts a fully unrolled radix-8 \ac{FFT} architecture. It employs single-precision floating-point arithmetic. Operating at 270MHz with a 0.9V power supply, the accelerator has a power consumption of 126 mW, making it a highly efficient solution for embedded signal processing delivering 16.5GFLOPS/W @ 270MHz.

Most existing hardware \ac{FFT} implementations either rely on fixed-point arithmetic~\cite{Pedram2014AHE}, which is unsuitable for frequency-domain processing in \ac{WUS} applications; or they do not target embedded applications. 
In this work, a memory-coupled multi-precision 16/32-bit floating-point radix-2 \ac{FFT} accelerator design is proposed. It is integrated as a Cluster-coupled \ac{HWPE}, where both the vector core and the accelerator share memory directly through the same interconnect. Unlike traditional \ac{FFT} accelerators, it does not require large internal buffers and instead relies entirely on the memory interconnect for data access. This approach improves the efficiency of the accelerators, reducing area usage while maintaining strong computing performance.
The accelerator supports up to 512 \ac{FFT} points for FP32 precision and 1024 \ac{FFT} points for FP16 precision, where both the real and imaginary components are represented in FP32 or FP16, respectively. 

\begin{figure*}[t!]
    \centering
    \includegraphics[width=1\linewidth]{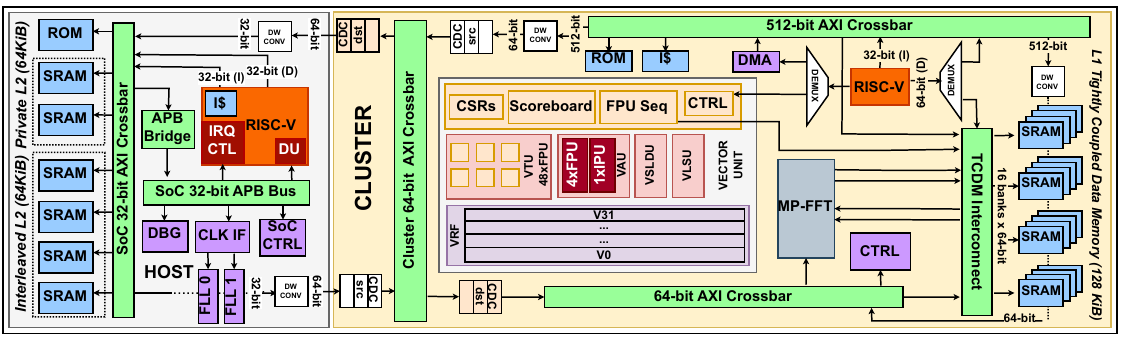}
    \caption{Maestro architecture with Host and Cluster domains. The Host domain includes a 32-bit RISC-V core, 64 KiB L2 memory, and two FLLs. The Cluster domain features a 32-bit RISC-V core, 128 KiB L1 memory, DMA, Vector Unit (VTU, VAU, VSLDU, VLSU, VRF), and MP-FFT accelerator, all sharing the L1 memory. }
    \label{fig:maestro_soc}
\end{figure*}

\subsection{Wearable Ultrasound Applications and Systems} \label{subsec:Wearable Ultrasound Applications and Systems}
State-of-the-art embedded systems for \ac{WUS} applications often rely on \acp{MCU} that do not play a central role in processing steps. Instead, they are used mainly to acquire echo data from the \ac{US} probes and transmit them to an external computer for further processing.

In~\cite{9872106}, a prosthetic hand control system employs four A-mode transducers and a STM32F7~\cite{STM32F7} \ac{MCU} to acquire echoes, perform basic filtering, and transmit data to a \ac{PC} for further feature extraction and classification. Although the MCU features a 32-bit Cortex-M7 running at 216MHz with a \ac{DSP} unit, it mainly serves for data acquisition, offloading signal processing---i.e., envelope extraction via \ac{HT}, \ac{PCA} for dimensionality reduction, and \ac{ML}-based classification---to an external computer.
Similarly, the wearable system in~\cite{10557691} uses a TMS320F28379D~\cite{TMS320F28379D} \ac{MCU} (dual-core, 200MHz) to digitize analog \ac{US} signals, apply a 1024-point \ac{FFT}, and transmit results to a \ac{PC}, where the \ac{LDA} model is applied for classification.

Xia et al.~\cite{8662605} propose a portable hybrid \ac{sEMG}/A-mode \ac{US} system for \ac{HMI}, integrating a composite sensor armband, signal acquisition modules, a \ac{DSP}, and a communication module.
The \ac{US} acquisition module consists of two main components: the Signal Excitation Part (SEP), responsible for generating and firing ultrasound waves via a beamformer, and the Signal Conditioning Part (SCP), which amplifies, filters, and optimizes the received echoes. The processed analog signals are digitized and further refined by a dsPIC33EP512MU814 \ac{DSP}~\cite{dsPIC33EP512MU814}, which applies filtering, amplification, and scaling to optimize signal quality for subsequent analysis.
%
This system has also been used by Zeng et al.~\cite{10208224}: in this work, the authors propose an adaptive \ac{CNN} based on A-mode \ac{US} for gesture recognition tasks. The raw A-mode \ac{US} data undergoes a series of preprocessing steps---\ac{TGC}, filtering, envelope detection via \ac{HT}, and log compression---before being fed into the \ac{CNN}.
Then, at test time, the feature extractor part is updated via backpropagation based on a pseudo-label. However, the test-time adaptation is still not performed locally but on an external \ac{PC}.

Frey et al. proposed an \ac{sEMG}-triggered Ultrasound System platform \cite{biowulpus}, which integrates \ac{sEMG} and \ac{US} for the long-term monitoring of muscle activity. 
This system integrates two state-of-the-art wearable devices: BioGAP\cite{biogap} for biosignal acquisition and WULPUS\cite{wulpus} for ultrasound data acquisition. It operates at 7.8mW when only BioGAP performs single-channel \ac{sEMG} measurement, with WULPUS in sleep mode, and consumes 29.8 mW when both \ac{sEMG} and \ac{US} systems are continuously active.
BioGAP is powered by GAP9, a commercial version of VEGA \cite{VEGA} along with an nRF52811 \cite{nRF52811} that provides data communication capability through a flexible \ac{BLE} communication. GAP9 is an ultra-low-power processor designed for hearable and battery-powered smart devices. It delivers 32.2 GMACs for \ac{ML} tasks and 15.6 GOPs for \ac{DSP} workloads, operating within a power envelope of up to 50mW.
WULPUS is built around an MSP430 \cite{MSP430} \ac{MCU} that manages communication with the \ac{US} transducers and forwards the incoming data to an nRF52832 \cite{nRF52832} \ac{MCU} that streams them via \ac{BLE} to an external \ac{PC} that receives both \ac{sEMG} and \ac{US} data from BioGAP and WULPUS, respectively. 

In contrast to these designs—which rely heavily on external computing or only feature limited \ac{DSP} capabilities—Maestro’s \ac{SoC} integrates a programmable vector unit, a hardware \ac{FFT} accelerator, and a tensor engine in a compact cluster, enabling high-performance frequency-domain analysis and \ac{ML}-based inference at the edge within a low power envelope. This unified and heterogeneous approach eliminates the need to transfer data across multiple components, allowing efficient, integrated embedded processing of \ac{WUS} signals.

%% file: 02.1-SoA_Table.tex
\begin{table*}[t]
\centering
\caption{Architectures for Vector, Tensor, FFT acceleration and Ultrasound Systems}
\label{tab:related-socs}
\begin{tabular}{|c|c|c|c|c|c|c|c|}
\hline
\multirow{3}{*}{\textbf{\begin{tabular}[c]{@{}c@{}}Architecture \end{tabular}}} &
  \multirow{3}{*}{\textbf{\begin{tabular}[c]{@{}c@{}}Technology \end{tabular}}} &
  \multirow{3}{*}{\textbf{\begin{tabular}[c]{@{}c@{}}CPU \end{tabular}}} &
  \multirow{3}{*}{\textbf{\begin{tabular}[c]{@{}c@{}}Max \\ Frequency\end{tabular}}} &
  \multirow{3}{*}{\textbf{\begin{tabular}[c]{@{}c@{}}Vector \\ Accelerator\end{tabular}}} &
  \multirow{3}{*}{\textbf{\begin{tabular}[c]{@{}c@{}}Tensor \\ Accelerator\end{tabular}}} &
  \multirow{3}{*}{\textbf{\begin{tabular}[c]{@{}c@{}}FFT \\ Accelerator\end{tabular}}} &
  \multirow{3}{*}{\textbf{\begin{tabular}[c]{@{}c@{}}Power \\ Envelope \end{tabular}}} \\
   &
   &
   &
   &
   &
   &
   &
   \\
   &
   &
   &
   &
   &
   &
   &
   \\ \hline
  \multirow{2}{*}{\begin{tabular}[c]{@{}c@{}}\textit{Ara \cite{ARA} } \end{tabular}} &
  \multirow{2}{*}{GF22 FDX} &
  \multirow{2}{*}{\begin{tabular}[c]{@{}c@{}}CVA6\\RV64GC\end{tabular}} &
  \multirow{2}{*}{1 GHz} &
  \multirow{2}{*}{\checkmark} &
  \multirow{2}{*}{-} &
  \multirow{2}{*}{-} &
  \multirow{2}{*}{\begin{tabular}[c]{@{}c@{}}763 mW\end{tabular}} \\
   &
   &
   &
   &
   &
   &
   &
   \\ \hline
  \multirow{2}{*}{\begin{tabular}[c]{@{}c@{}}\textit{YUN \cite{YUN} } \end{tabular}} &
  \multirow{2}{*}{TSMC 65} &
  \multirow{2}{*}{\begin{tabular}[c]{@{}c@{}}CVA6\\RV64GC\end{tabular}} &
  \multirow{2}{*}{280 MHz} &
  \multirow{2}{*}{\checkmark} &
  \multirow{2}{*}{-} &
  \multirow{2}{*}{-} &
  \multirow{2}{*}{\begin{tabular}[c]{@{}c@{}}330 mW\end{tabular}} \\
   &
   &
   &
   &
   &
   &
   &
   \\
   \hline
 \multirow{2}{*}{\begin{tabular}[c]{@{}c@{}}\textit{Speed \cite{SPEED}} \end{tabular}} &
  \multirow{2}{*}{TSMC 28} &
  \multirow{2}{*}{\begin{tabular}[c]{@{}c@{}}CVA6\\RV64GC\end{tabular}} &
  \multirow{2}{*}{1.05 GHz} &
  \multirow{2}{*}{\checkmark} &
  \multirow{2}{*}{\checkmark} &
  \multirow{2}{*}{-} &
  \multirow{2}{*}{\begin{tabular}[c]{@{}c@{}}533 mW\end{tabular}} \\
   &
   &
   &
   &
   &
   &
   & \\
   \hline
 \multirow{2}{*}{\begin{tabular}[c]{@{}c@{}}\textit{Spatz \cite{Cavalcante2023SpatzCC}} \end{tabular}} &
  \multirow{2}{*}{GF 12LPP} &
  \multirow{2}{*}{\begin{tabular}[c]{@{}c@{}}Snitch\\ RV32IMC\end{tabular}} &
  \multirow{2}{*}{1 GHz} &
  \multirow{2}{*}{\checkmark} &
  \multirow{2}{*}{-} &
  \multirow{2}{*}{-} &
  \multirow{2}{*}{\begin{tabular}[c]{@{}c@{}}164 mW\end{tabular}} \\
   &
   &
   &
   &
   &
   &
   &
   \\
   \hline
 \multirow{2}{*}{\begin{tabular}[c]{@{}c@{}}\textit{Gemmini \cite{gemmini}} \end{tabular}} &
  \multirow{2}{*}{GF22 FDX} &
  \multirow{2}{*}{\begin{tabular}[c]{@{}c@{}}Rocket\\ RV64GC\end{tabular}} &
  \multirow{2}{*}{961 MHz} &
  \multirow{2}{*}{-} &
  \multirow{2}{*}{\checkmark} &
  \multirow{2}{*}{-} &
  \multirow{2}{*}{\begin{tabular}[c]{@{}c@{}}-\end{tabular}} \\
   &
   &
   &
   &
   &
   &
   &
   \\ \hline
 \multirow{2}{*}{\begin{tabular}[c]{@{}c@{}}\textit{Tsunami \cite{Tsunami}} \end{tabular}} &
  \multirow{2}{*}{65nm CMOS} &
  \multirow{2}{*}{-} &
  \multirow{2}{*}{200 MHz} &
  \multirow{2}{*}{-} &
  \multirow{2}{*}{\checkmark} &
  \multirow{2}{*}{-} &
  \multirow{2}{*}{\begin{tabular}[c]{@{}c@{}}419 mW\end{tabular}} \\
   &
   &
   &
   &
   &
   &
   &
   \\ \hline
 \multirow{2}{*}{\begin{tabular}[c]{@{}c@{}}\textit{DARKSIDE \cite{Garofalo2023DARKSIDEAH}} \end{tabular}} &
  \multirow{2}{*}{TSMC 65} &
  \multirow{2}{*}{\begin{tabular}[c]{@{}c@{}}8xRISCY-NN\\RVC32IMFXpulpNN2\end{tabular}} &
  \multirow{2}{*}{290 MHz} &
  \multirow{2}{*}{-} &
  \multirow{2}{*}{\checkmark} &
  \multirow{2}{*}{-} &
  \multirow{2}{*}{\begin{tabular}[c]{@{}c@{}}213 mW\end{tabular}} \\
   &
   &
   &
   &
   &
   &
   &
   \\ \hline
 \multirow{2}{*}{\begin{tabular}[c]{@{}c@{}}\textit{ECHOES \cite{Echoes}} \end{tabular}} &
  \multirow{2}{*}{TSMC 65} &
  \multirow{2}{*}{\begin{tabular}[c]{@{}c@{}}CV32e40p\end{tabular}} &
  \multirow{2}{*}{350 MHz} &
  \multirow{2}{*}{-} &
  \multirow{2}{*}{-} &
  \multirow{2}{*}{\checkmark} &
  \multirow{2}{*}{\begin{tabular}[c]{@{}c@{}}134 mW\end{tabular}} \\
   &
   &
   &
   &
   &
   &
   &
   \\ \hline
 \multirow{2}{*}{\begin{tabular}[c]{@{}c@{}}\textit{\cite{8240277}} \end{tabular}} &
  \multirow{2}{*}{16nm FinFET} &
  \multirow{2}{*}{\begin{tabular}[c]{@{}c@{}}Rocket-Chip\end{tabular}} &
  \multirow{2}{*}{940 MHz} &
  \multirow{2}{*}{-} &
  \multirow{2}{*}{-} &
  \multirow{2}{*}{\checkmark} &
  \multirow{2}{*}{\begin{tabular}[c]{@{}c@{}}22.6 mW\end{tabular}} \\
   &
   &
   &
   &
   &
   &
   &
   \\ \hline
 \multirow{2}{*}{\begin{tabular}[c]{@{}c@{}}\textit{\cite{10509999}} \end{tabular}} &
  \multirow{2}{*}{65nm CMOS} &
  \multirow{2}{*}{-} &
  \multirow{2}{*}{100 MHz} &
  \multirow{2}{*}{-} &
  \multirow{2}{*}{-} &
  \multirow{2}{*}{\checkmark} &
  \multirow{2}{*}{\begin{tabular}[c]{@{}c@{}}59.11 mW\end{tabular}} \\
   &
   &
   &
   &
   &
   &
   &
   \\ \hline
 \multirow{2}{*}{\begin{tabular}[c]{@{}c@{}}\textit{\cite{9926333}} \end{tabular}} &
  \multirow{2}{*}{TSMC 28} &
  \multirow{2}{*}{-} &
  \multirow{2}{*}{270 MHz} &
  \multirow{2}{*}{-} &
  \multirow{2}{*}{-} &
  \multirow{2}{*}{\checkmark} &
  \multirow{2}{*}{\begin{tabular}[c]{@{}c@{}}126 mW\end{tabular}} \\
   &
   &
   &
   &
   &
   &
   &
   \\ \hline
 \multirow{2}{*}{\begin{tabular}[c]{@{}c@{}}\textit{STM32F7 \cite{STM32F7}} \\ \textit{(product)} \end{tabular}} &
  \multirow{2}{*}{90nm CMOS} &
  \multirow{2}{*}{\begin{tabular}[c]{@{}c@{}}Cortex-M7 + \\ DSP \end{tabular}} &
  \multirow{2}{*}{216 MHz} &
  \multirow{2}{*}{-} &
  \multirow{2}{*}{-} &
  \multirow{2}{*}{-} &
  \multirow{2}{*}{\begin{tabular}[c]{@{}c@{}}900 mW\end{tabular}} \\
   &
   &
   &
   &
   &
   &
   &
   \\ \hline
   \multirow{2}{*}{\begin{tabular}[c]{@{}c@{}}\textit{TMS320F28379D \cite{TMS320F28379D}} \\ \textit{(product)} \end{tabular}} &
  \multirow{2}{*}{-} &
  \multirow{2}{*}{\begin{tabular}[c]{@{}c@{}}2xTMS320C28x +\\ TMU + VCU-II \end{tabular}} &
  \multirow{2}{*}{200 MHz} &
  \multirow{2}{*}{-} &
  \multirow{2}{*}{-} &
  \multirow{2}{*}{\checkmark} &
  \multirow{2}{*}{\begin{tabular}[c]{@{}c@{}}576 mW\end{tabular}} \\
   &
   &
   &
   &
   &
   &
   &
   \\ \hline
   \multirow{2}{*}{\begin{tabular}[c]{@{}c@{}}\textit{dsPIC33EP512MU814 \cite{dsPIC33EP512MU814}} \\ \textit{(product)} \end{tabular}} &
  \multirow{2}{*}{-} &
  \multirow{2}{*}{\begin{tabular}[c]{@{}c@{}}16-bit DSP \end{tabular}} &
  \multirow{2}{*}{70 MHz} &
  \multirow{2}{*}{-} &
  \multirow{2}{*}{-} &
  \multirow{2}{*}{-} &
  \multirow{2}{*}{\begin{tabular}[c]{@{}c@{}}350 mW \end{tabular}} \\
   &
   &
   &
   &
   &
   &
   &
   \\ \hline
   \multirow{2}{*}{\begin{tabular}[c]{@{}c@{}}\textit{GAP9 /} \\ \textit{VEGA \cite{VEGA} } \end{tabular}} &
  \multirow{2}{*}{GF22 FDX} &
  \multirow{2}{*}{\begin{tabular}[c]{@{}c@{}} 10 $\times$RI5CY \\ RVC32IMFXpulp + SF\end{tabular}} &
  \multirow{2}{*}{370 MHz} &
  \multirow{2}{*}{-} &
  \multirow{2}{*}{\checkmark} &
  \multirow{2}{*}{-} &
  \multirow{2}{*}{\begin{tabular}[c]{@{}c@{}}50 mW \end{tabular}} \\
   &
   &
   &
   &
   &
   &
   &
   \\ \hline
 \multirow{2}{*}{\begin{tabular}[c]{@{}c@{}}\textit{This work} \end{tabular}} &
  \multirow{2}{*}{TSMC 65} &
  \multirow{2}{*}{\begin{tabular}[c]{@{}c@{}}Snitch\\ RV32IMC\end{tabular}} &
  \multirow{2}{*}{210 MHz} &
  \multirow{2}{*}{\checkmark} &
  \multirow{2}{*}{\checkmark} &
  \multirow{2}{*}{\checkmark} &
  \multirow{2}{*}{\begin{tabular}[c]{@{}c@{}}212 mW\end{tabular}} \\
   &
   &
   &
   &
   &
   &
   &
   \\ \hline
\end{tabular}
\end{table*}

%% file: 03-Maestro_Architecture.tex
\section{Maestro Architecture} \label{sec:Maestro Architecture}
The Maestro \ac{SoC}, comprises two distinct clock domains, as depicted in Fig. \ref{fig:maestro_soc}: (i) the Host, featuring the main 32-bit core of the \ac{SoC}, and (ii) the Cluster, equipped with an advanced \ac{VU} extended with a \ac{TU} and a 16/32-bit multi-precision \ac{FFT} processing engine.

\subsection{Host Domain} \label{subsec:Host Domain}
The Host domain features a compact, 2-stage, 32-bit RISC-V RV32IMC core \cite{8106976}, optimized for low-cost and energy-efficient operation. This core is supported by a memory subsystem consisting of 64KiB of shared L2 \ac{TCDM} organized into 4 interleaved SRAM banks, four additional 16KiB interleaved SRAM banks designated as private memory, and a boot ROM. All components are interconnected through a 32-bit \ac{AXI}-4 crossbar, ensuring efficient access and integration. Additionally, an \ac{APB} interconnect provides access to \ac{SoC} control registers and a debug interface, enabling system management and debugging capabilities. To ensure robust clocking for both the Host and Cluster domains, the architecture incorporates two dedicated \ac{FLLs}.
This setup showcases the adaptability of the Cluster, which can be incorporated with alternative Host subsystems to meet the needs of diverse applications or architectural requirements.

\subsection{Cluster Domain} \label{subsec:Cluster Domain}
The Cluster domain hosts the core contribution of this work, featuring a single-stage RISC-V RV32IMAFD Snitch scalar core \cite{Snitch}, augmented with an RVV1.0 Zve64d-compliant Spatz \ac{VU} \cite{Cavalcante2023SpatzCC}. This architecture enhances the \ac{VU} with a fully programmable \ac{TU} based on RedMulE architecture \cite{TORTORELLA2023122}, forming a unified and versatile area-optimized \ac{VTU} for high-performance and energy efficiency vector-tensor operations. 
Moreover, the Cluster includes a floating-point multi-precision 16/32-bit \ac{FFT} accelerator, integrated as \ac{HWPE} sharing the memory with the Cluster through a low-latency L1 \ac{TCDM} interconnect.

The memory subsystem features a shared 128KiB L1 \ac{TCDM} organized into 16 interleaved SRAM banks to optimize parallel access and minimize contention. Additionally, a 512-bit/cycle read and 512-bit/cycle write \ac{DMA} controller \cite{10.1109/TC.2023.3329930} facilitates data movement between L2 and L1.
The interconnect of the Cluster is designed to ensure efficient communication both within its internal components and with external systems. A 64-bit \ac{AXI}-4 crossbar serves as the primary interface between the Cluster and the Host, enabling bidirectional data transfer. At the Cluster boundary, \ac{CDC} First-In First-Out (FIFO) queues to ensure communication with the host domain enabling the two blocks to work at different frequencies. 
The Cluster features a hierarchical interconnect designed to handle efficient data and instruction transfers. At the top level, a high-throughput 512-bit wide \ac{AXI}-4 crossbar enables data transfers, reducing the number of cycles required for large transactions. This interconnect links the instruction cache and the boot ROM to the scalar core, while the \ac{DMA} controller facilitates data movement between L1 and L2 memory. A smaller 64-bit AXI-4 interconnect provides access to the Cluster’s control registers and enables communication with the \ac{FFT} accelerator.

The scalar core pre-decodes and dispatches vector instructions via a generic accelerator interface, which also facilitates communication with the \ac{DMA}, allowing parallel execution with the \ac{VU}. The \ac{VU} features 32$\times$512-bit latch-based vector registers coupled with four \ac{FUs}, featuring a \ac{VLEN} of 512-bit and maintaining a throughput of 64-bit/cycle. The \ac{VAU} consists of four mixed-precision \ac{FPUs} that support FP64, FP32, FP16, BF16, and FP8 (E4M3, E5M2), along with dot product operations featuring higher-precision accumulation, where 8-bit and 16-bit inputs are accumulated into 16-bit and 32-bit results (G16-32/G8-16). Additionally, it includes an \ac{IPU} that supports 8-bit, 16-bit, and 32-bit integer operations. The \ac{VSLDU} handles vector permutation operations, including vector slide up/down and vector move instructions. The \ac{VLSU} manages the memory-to-register and register-to-memory transfer, supporting indexed, non-unit-strided, and unit-strided memory accesses. Along these three \ac{FUs} the \ac{VTU} described in the following Section enhances tensor operations.

\input{03.1-Vector_Tensor}

\input{03.2-Floating-Point-FFT}

%% file: 03.1-Vector_Tensor.tex
\section{Unified Vector-Tensor Unit} \label{sec:Unified Vector Tensor Unit}
\begin{figure}[t]
    \centering
    \includegraphics[width=1\linewidth]{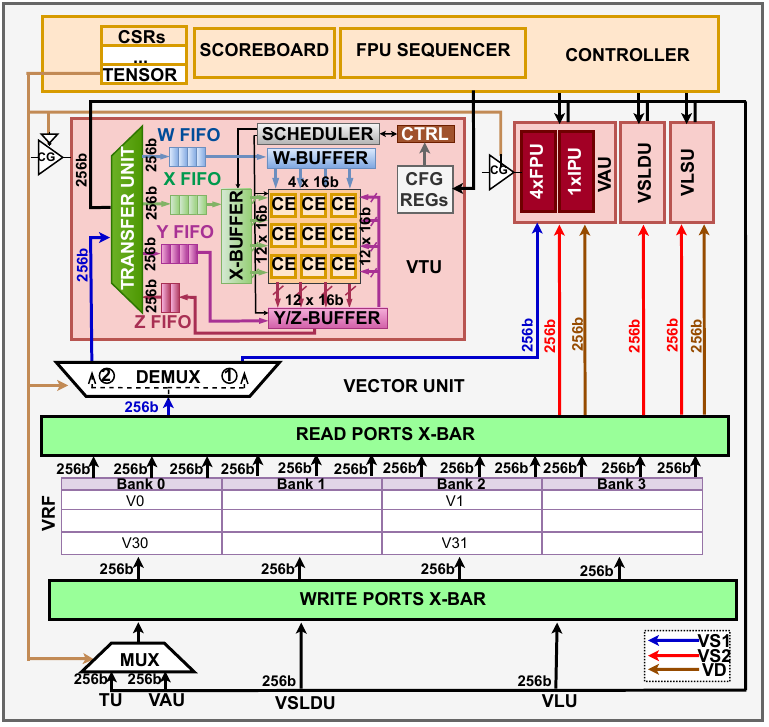}
    \caption{Vector-Tensor Unit (VTU) architecture and Vector Register File (VRF) interconnect with Vector Functional Units (VFU). The VTU features the transfer unit and X, W, and Z/W buffers filling the 4$\times$12 Computing Elements (CEs) datapath. The Tensor Control Status Register (TCSR) manages Vector-Tensor capabilities.}
    \label{fig:cluster_vtu}
\end{figure}

In this Section, we focus specifically on the combination of the \ac{VU} and the \ac{VTU}, which is depicted in detail in Fig. \ref{fig:cluster_vtu}. To realize the \ac{VTU}, we augmented the Spatz's \ac{FUs} with a tightly integrated \ac{TU} \cite{TORTORELLA2023122} for \ac{GEMM} and \ac{GEMM-Ops} execution consisting of two-dimensional array of 12$\times$4 \ac{CEs}, resulting in a peak performance rate of 48 \ac{FMA}/cycle.

\begin{figure*}[t!]
    \centering
    \includegraphics[width=1\linewidth]{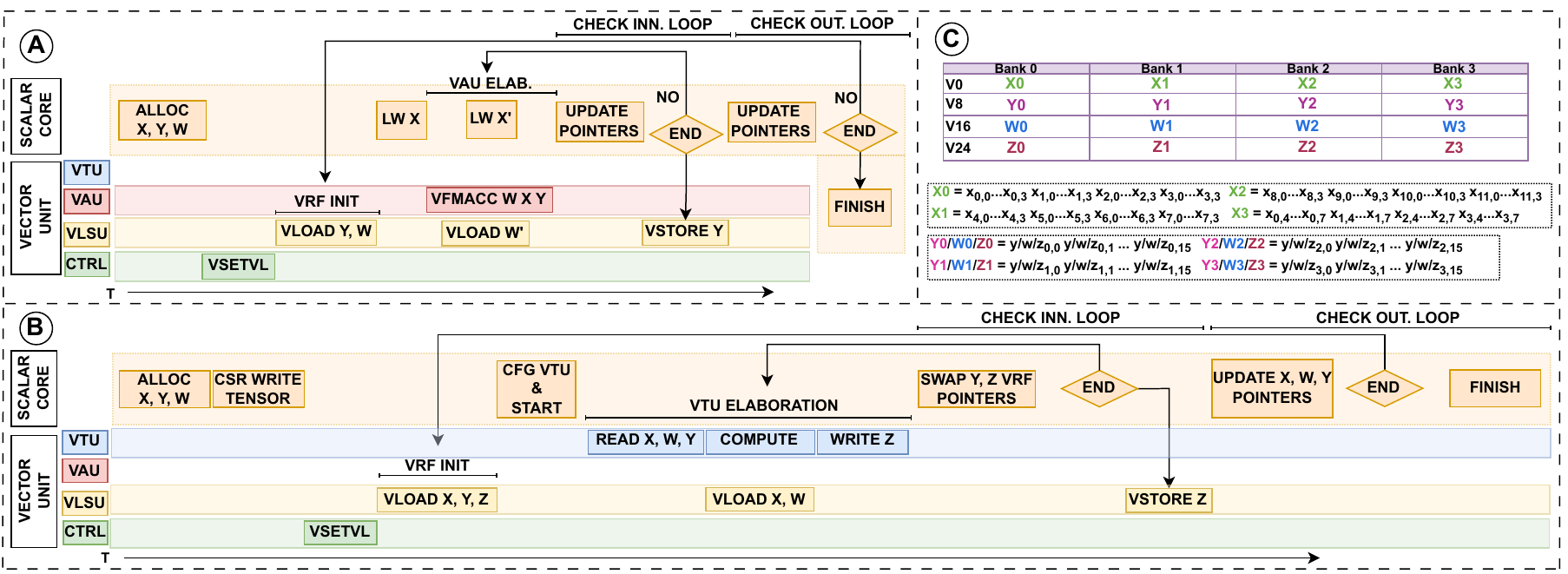}
    \caption{Vector Arithmetic Unit (VAU) and Vector-Tensor Unit (VTU) temporal execution flow of a general matrix-matrix multiplication (A-B). Vector Register File (VRF) data organization when executing on VTU (C).}
    \label{fig:vu_vtu_execution_flow_and_VRF_VTU}
\end{figure*}

\subsection{Vector-Tensor Unit Architecture} \label{subsec:Vector-Tensor Unit Architecture}

A Transfer Unit handles data transfers to and from the \ac{VRF}, providing an input and output bandwidth of 256-bits/cycle. It is equipped with three dedicated buffers: the X-Buffer, which updates the 12-column inputs of the X matrix; the W-Buffer, composed of 4 shift registers, which efficiently distributes new 64-bit/cycle inputs of the W matrix to the \ac{CEs}; and the Z-Buffer, which simultaneously stores and forwards computed outputs while pre-loading elements from the Y input matrix, functioning as both an output buffer and a pre-load buffer.

The \ac{VRF} is equipped with 6$\times$256-bit-wide read ports and 3$\times$256-bit-wide write ports to support high-bandwidth vector operations efficiently. Of the six read ports, three are dedicated to the \ac{VAU}, providing the 3×256-bits/cycle input bandwidth required for operations such as \ac{VFMACC}, which processes three vector inputs. The \ac{VAU} also utilizes one write port to handle its 256-bits/cycle output. Two additional read ports are allocated to the \ac{VLSU} for accessing vector data, while the remaining read port is assigned to the \ac{VSLDU}, supporting sliding operation. The \ac{VRF}, organized into four banks, features a multi-ported architecture with 3$\times$256-bit wide read ports and 1$\times$256-bit wide write port per bank.

To efficiently arbitrate accesses among competing functional units, a priority-based scheme is applied both for read and write operations on each bank. For read operations, port 0 is primarily dedicated to the \ac{VAU}, granting it the highest priority for accessing the VS2 operand. If the \ac{VAU} does not utilize the port in a given cycle, the \ac{VLSU} is granted access to retrieve the VS2 operand. Similarly, port 1 follows a hierarchical priority scheme, with the \ac{VAU} having precedence for accessing the VS1 operand. When the \ac{VAU} does not require the port, the \ac{VSLDU} gains access to fetch its VS2 operand. Finally, port 2 prioritizes the \ac{VAU} for reading the VD operand. In the absence of \ac{VAU} activity, the \ac{VLSU} is permitted to use the port for VD operand handling. 

The write priority scheme for the single write port of each \ac{VRF} bank ensures efficient operation by giving the highest priority to the \ac{VAU} for writing arithmetic results. If unused by the \ac{VAU}, the \ac{VLSU} gains access for memory operations, followed by the \ac{VSLDU}, which writes only when neither the \ac{VAU} nor \ac{VLSU} requires the port. 

A \ac{TCSR} manages the clock-gating mechanism to disable the unused \ac{FUs}, optimizing power efficiency while enabling tensor capabilities by multiplexing the \ac{VAU}'s VS1 read port to the \ac{TU}. Additionally, the \ac{TU} and the \ac{VAU} share the same write port to the \ac{VRF}.

\subsection{Buffer Reduction} \label{subsec:Buffer Reduction}

A key aspect of this integration is the \ac{VTU}'s buffer size reduction, which is made possible by leveraging the \ac{VRF} as a shared resource between vector and tensor operations. 
Using a single 256-bit read port from the \ac{VRF}, the design supplies data for all 48 \ac{CEs} over three clock cycles. Each \ac{CE} uses its X value for 16 clock cycles, eliminating the need for repeated reads. However, due to the staggered timing of computations across columns—immediate for the first column and delayed by 4, 8, and 12 clock cycles for the subsequent columns—an additional register is introduced for each \ac{CE} to hold the new X value until it becomes valid for the respective column. 
This dual-register approach keeps operations synchronized and ensures they run correctly within the required timing.
The result is a more efficient X-buffer design that maintains performance with significantly reduced memory overhead.

In the original \ac{TU}, each \ac{CE} has a queue of 4×16-bit elements, along with two additional queue slots to synchronize data usage across the processing array, leading to a total buffer size of 576 bytes, calculated as (\ref{full_x_equation}):

\begin{equation}
\label{full_x_equation}
(2+4)queue*48CEs*16bits = 576 B
\end{equation}
In the integrated \ac{VTU} design, both queue sizes have been halved per each \ac{CE}, requiring only 288 Bytes (\ref{halved_x_equation}).
\begin{equation}
\label{halved_x_equation}
(1+2)queue*48CEs*16bits = 288 B
\end{equation}

Table \ref{tab:redmule_buffer_capacity} and \ref{tab:redmule_area} provide a detailed comparison of the original \ac{TU} and the integrated \ac{VTU}, demonstrating that this optimization results in a 26.5\% reduction of the whole buffer capacity with consequent 19.8\% buffer area saving.  Given that buffers account for 25.6\% of the total RedMule area~\cite{TORTORELLA2023122}, this optimization reduces the \ac{TU} area by 6\%.

\begin{table}[h!]
 \caption{Tensor Unit Buffers Capacity}
 \label{tab:redmule_buffer_capacity}
 \centering
 \begin{tabular}{@{}lccr@{}} \toprule
  \textbf{} & \textbf{RedMule\cite{TORTORELLA2023122}} & \textbf{This work} & \textbf{Reduction} \\ \midrule
  X buffer  & 576 B & 288 B & -50\% \\
  Y buffer  & 384 B & 384 B &  0\%  \\
  W buffer  & 128 B & 128 B &  0\%  \\
  \hline
  Total     & 1088 B & 800 B &  -26.5\% \\ \bottomrule
 \end{tabular}
\end{table}

\begin{table}[h!]
 \caption{Tensor Unit Area Breakdown TSMC65 Technology}
 \label{tab:redmule_area}
 \centering\begin{tabular}{@{}lccr@{}} \toprule
  \textbf{} & \textbf{RedMule\cite{TORTORELLA2023122}} & \textbf{This work} & \textbf{Area Change}
  \\ \textbf{} & \textbf{[mm$^2$]} & \textbf{[mm$^2$]} & \textbf{[\%]} \\ \midrule
  Engine   & 0.572  & 0.567  &  -0.9  \\
  Buffers  & 0.241  & 0.194  &  -19.8 \\
  Transfer Unit & 0.077  & 0.074  &  -3.1  \\
  Scheduler& 0.026  & 0.026  &  -1.8  \\
  Control  & 0.025  & 0.025  &  -1.7  \\
  \hline
  Total Area & 0.942  & 0.886  &  -6.0 \\ \bottomrule
 \end{tabular}
\end{table}

\subsection{VAU and VTU Execution Flow} \label{subsec:VU and VTU Execution Flow}

Fig. \ref{fig:vu_vtu_execution_flow_and_VRF_VTU} depicts the execution flow of a \ac{GEMM}, emphasizing how the two computational units—the \ac{VAU} (A) and the \ac{VTU} (B)—are engaged to perform the operation. 
When executing on the \ac{VAU} Fig. \ref{fig:vu_vtu_execution_flow_and_VRF_VTU}a, the scalar core allocates input and output matrices, then fetches and pre-decodes the vector instructions, delegating their execution to the \ac{VFUs}. At the inner computation loop setup, the \ac{VU} dynamically sets the vector length and loads the W and X data into the \ac{VRF}.
We adopt an outer product algorithm to maximize parallelism: a scalar from the input tensor is multiplied by a vector from the weight tensor while the Snitch and the \ac{VLSU} pre-load, respectively, scalar and vector inputs for the next round. At the end of each iteration, the partial results are stored back in the memory, and the scalar core reassesses the computation's status, updating the pointers as needed.
This execution flow relies on tight cooperation between the scalar core and the \ac{VU}, as the scalar core handles the continuous pre-decoding, fetching, and dispatching of vector instructions. 

When performing elaboration on the \ac{VTU} Fig. \ref{fig:vu_vtu_execution_flow_and_VRF_VTU}b, the scalar core handles the initialization phase, which involves allocating the matrices and enabling tensor capabilities by configuring the \ac{TCSR}, as described in Subsection \ref{subsec:Vector-Tensor Unit Architecture}. Following this, the scalar core delegates the task of loading X, W, and Y data from the L1 memory into the \ac{VRF}, leveraging a \ac{LMUL}=8 configuration. This approach combines eight vector registers into a larger logical register, optimizing data locality and access efficiency.

Fig. \ref{fig:vu_vtu_execution_flow_and_VRF_VTU}c depicts how data are organized into the \ac{VRF} during this process. Specifically, the V0 register is used for managing X-data, with its content distributed across four memory banks. Each bank holds 16$\times$16-bit elements, sufficient to populate four rows of 4$\times$\ac{CEs} progressively. For example, X0 stores elements for CE0,0 to CE3,3, X1 stores elements for CE4,0 to CE7,3, and so on, ensuring efficient data alignment and distribution across the computation elements. The V8 and V16 registers are designated for managing Y and W data, respectively. As illustrated in the legend, these registers follow a consistent loading pattern, with each memory bank handling 16$\times$16-bit elements required to refill the Y and W buffers efficiently. The \ac{VTU} uses the V24 register to store Z elements.

When the \ac{VRF} initialization is complete, the scalar core configures and triggers the \ac{VTU} execution, while the \ac{VLSU} simultaneously loads new X and W values. Upon completion, the \ac{VTU} streams the computed Z back to the \ac{VRF}. The scalar core then applies double-buffering, swapping the Y and Z pointers used by the \ac{VTU} while relaunching the inner loop computation. Once this process concludes, the \ac{VLSU} stores the final Z output, and the outer loop check is performed by updating pointers and reinitializing the \ac{VRF} for the next iteration.

%% file: 03.2-Floating-Point-FFT.tex
\section{Mixed precision FFT accelerator} \label{sec:Mixed precision FFT Accelerator}
\begin{figure}[t]
    \centering
    \includegraphics[width=1\linewidth]{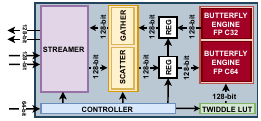}
    \caption{Multi-Precision FFT Accelerator featuring the Streamer, Controller, Gather and Scatter units, along with two Butterfly Engines: one C32 and one C64.}
    \label{fig:mp_fft}
\end{figure}

The MP-FFT accelerator shown in Fig. \ref{fig:mp_fft} implements a configurable mixed-precision floating-point FP16 and FP32 radix-2 Cooley-Tukey \ac{FFT} \cite{Cooley1965AnAF} following the template of the fixed-precision accelerator in \cite{bertaccini2021buffer}.
In this Section, we descibe in detail the architecture of the MP-FFT, as well as the novel floating-point butterfly implementation that we propose and the integration in the Maestro system.

\subsection{MP-FFT Architecture} \label{subsec:MP-FFT Architecture}
The accelerator targets IEEE-754 single- and half-precision floating-point arithmetic \cite{8766229} and supports up to 1024-point and 512-point \ac{FFT} computations for complex data with 16-bit and 32-bit real and imaginary parts, respectively, defined as C32 and C64 formats. 
Its design implementation consists of several submodules organized in a pipelined architecture described as follows.

The Streamer serves as a specialized \ac{DMA} unit, efficiently managing data transfers between L1 memory and the accelerator. The MP-FFT accelerator includes four 64-bit memory ports, divided into two input and two output ports. The Controller is responsible for managing the status of various accelerator modules, coordinating their operations, handling the register file used for programming, and controlling the peripheral interconnect.
The Gather-Scatter Unit reorganizes the samples. The Scatter module inserts new data from the Streamer into the butterfly registers, while the Gather module extracts processed data and returns it to the Streamer to store it in memory. The Butterfly Registers, comprising a pair of four 64-bit registers, act as temporary storage positioned in front of the Butterfly Unit, which is the main computation engine of the MP-FFT.
For coefficient management, the Twiddle Factor \ac{LUTs} store precomputed twiddle factors corresponding to the maximum \ac{FFT} points. The architecture integrates a C64 LUT with 65 elements and two C32 \ac{LUTs}, each containing 129 elements, ensuring efficient multi-precision support. 

At the core of the accelerator, the Floating-Point Butterfly Unit performs the radix-2 \ac{FFT} computation. It consists of two Butterfly Engines, one FP16 for C32 precision and one FP32 for C64 precision. Each Butterfly Engine incorporates two fused-arithmetic Dual-Output \ac{SDOTP} modules, which are discussed in detail in Sec. \ref{subsec:Floating-Point Butterfly Engine Implementation}.
Since the accelerator reuses larger engines to compute lower-precision butterflies, the accelerator can compute either one C64 or two C32 butterflies at each cycle. 
In case an \ac{FFT} C32 is performed, the inputs are rearranged to words with the right data size and are scattered to both butterfly engines reusing the C64 Butterfly for C32 computation. At the output, the results of the butterflies are concatenated to come back to two 32-bit complex words.

\subsection{Floating-Point Butterfly Engine Implementation} \label{subsec:Floating-Point Butterfly Engine Implementation}
Each MP-\ac{FFT}'s Butterfly Engine implements the radix-2 \ac{DIT} butterfly operation. 

\mycomment{
\begin{figure}[t]
    \centering
    \includegraphics[width=1\linewidth]{Figures/radix2_structure.pdf}
    \caption{Radix-2 DIT butterfly.}
    \label{fig:radix2_structure}
\end{figure}
}

By exploiting its symmetry, the butterfly can be rewritten so that the real and imaginary parts of both the left and right-sided can be expressed in the form:
\begin{equation}
\label{eq:vediamo}
E \pm (A\cdot B\pm C\cdot D)
\end{equation}

This operation can be conveniently implemented as a high-accuracy fused-arithmetic floating-point instruction.

Exploiting this observation, we designed each Butterfly Engine with two fused-arithmetic Dual-Output Sum of Dot Products (DO-SDOTP) units.
DO-SDOTP handles subnormal numbers consistently with other IEEE-754 operations, and consists of a five-operand module: four inputs and an accumulator input/output. 

\begin{figure}[t]
    \centering
    \includegraphics[width=1\linewidth]{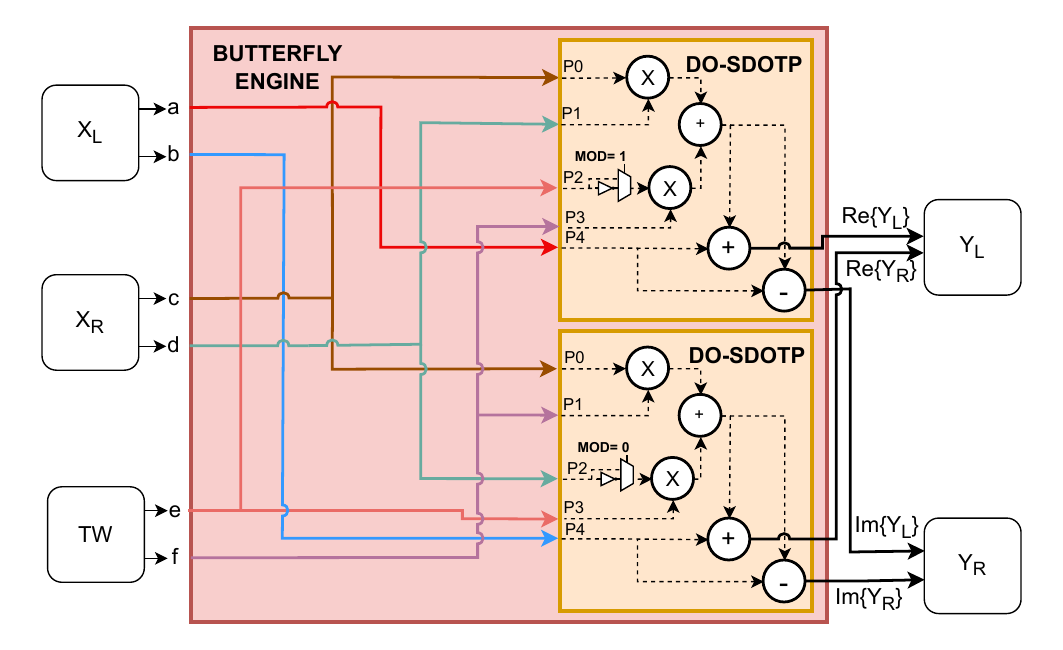}
    \caption{Floating-Point Radix-2 Butterfly Engine.}
    \label{fig:butterfly_engine}
\end{figure}

We designed the DO-SDOTP module as a parametric design that can be configured to support multiple precision formats ranging from FP8 to FP64; we employed the FP16 and FP32 versions in Maestro. It includes a special input signal, MOD, which inverts the sign of the third operand. 
As a result, the implemented hardware module is illustrated in Fig. \ref{fig:butterfly_engine}. 

Compared to a conventional Radix-2 FP datapath, the fused DO-SDOTP datapath avoids precision loss in low-bitwidth FP formats by reducing rounding errors. 
Each unit instance operates within the largest exponent and mantissa widths defined by the format parameterization, allowing lower-precision computations to be mapped onto the same datapath by assigning narrower exponent and mantissa fields to specific bit positions.
The computation starts with mantissa multiplication, producing values at double precision. These are sorted, zero-padded, and aligned by right-shifting based on the exponent difference. A second sorting stage incorporates a fifth operand, ensuring precision consistency. Two adders then perform addition and subtraction, with the sign bit inverted in the exponent datapath. Finally, results undergo normalization and rounding, restoring the original precision.

\begin{figure}[t!]
    \centering
    \includegraphics[width=1\linewidth]{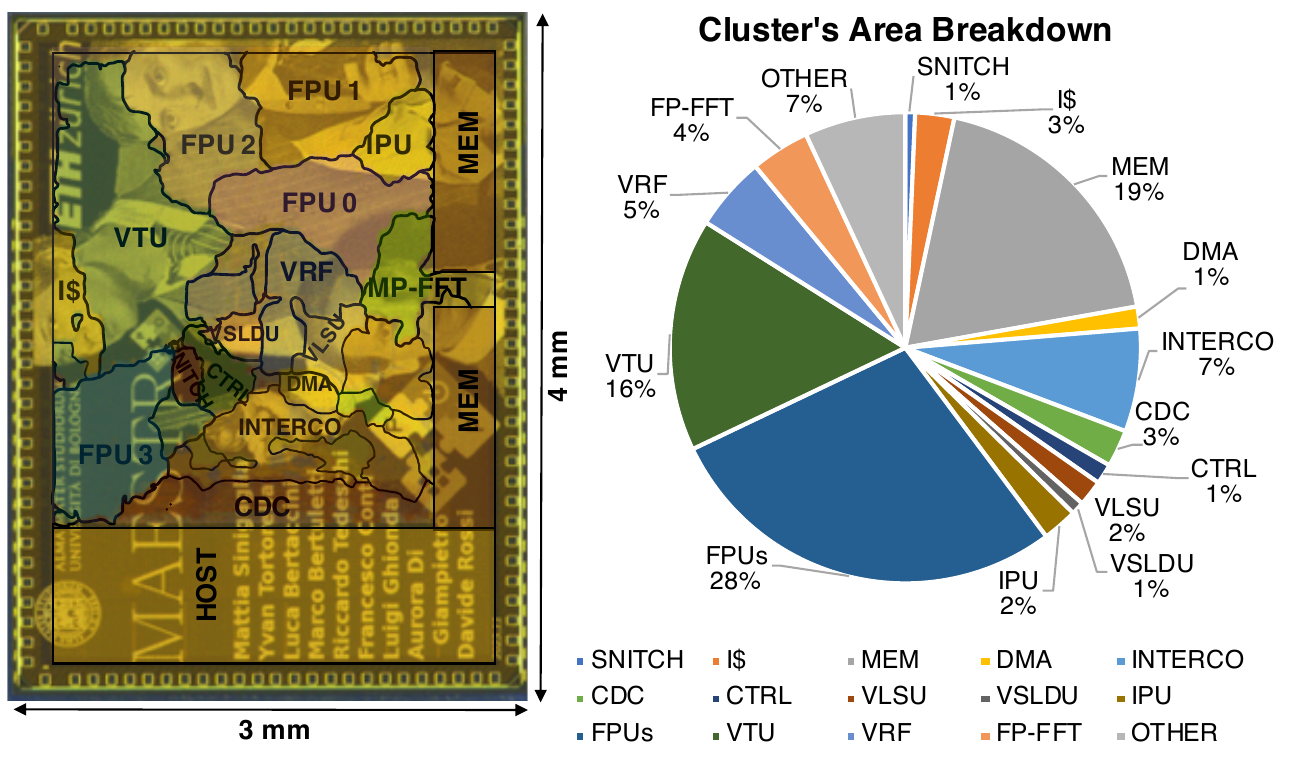}
    \caption{Chip micrograph and Cluster's area breakdown.}
    \label{fig:micrograph}
\end{figure}

\begin{figure*}[t!]
    \centering
    \includegraphics[width=1\linewidth]{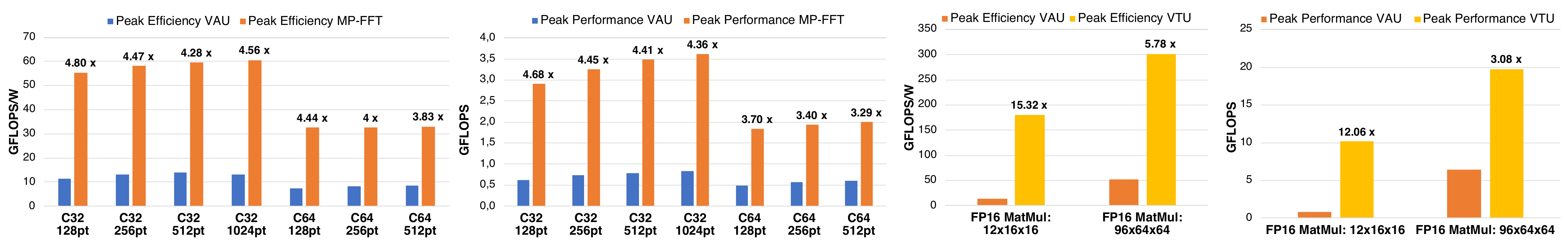}
    \caption{Peak Performance and Energy Efficiency for FFT and MatMul Kernels over VAU, VTU, and MP-FFT accelerator.}
    \label{fig:kernel_perf}
\end{figure*}

\subsection{FFT Accelerator Integration} \label{subsec:FFT Accelerator Integration}
The \ac{FFT} accelerator is integrated into Maestro’s cluster as an \ac{HWPE} engine, tightly coupled with the 128KiB L1 memory through a low-latency interconnect.  
The programmability of the accelerator is ensured by connecting the \ac{HWPE}’s control port to the 64-bit AXI-4 crossbar. 
When an \ac{FFT} computation needs to be accelerated, the Scalar core in the system configures the accelerator through its memory-mapped control registers and triggers the execution of a job.

The accelerator begins by loading a set of left-wing samples from consecutive memory addresses through the Streamer. In the subsequent cycle, it recovers the corresponding right-wing samples for the same butterflies. The Scatter Unit then reorganizes these samples into two of the four 64-bit registers positioned in front of the Butterfly Unit. Once both the left and right wings of the butterflies are stored in the registers, the Butterfly Unit processes them. 
After computation, the processed partial results are sampled back into the registers.
As the computation progresses, the Scatter Unit continues to insert new inputs into the registers, overwriting samples that the Butterfly Unit has already processed. Meanwhile, the Gather Unit works in tandem with the Scatter Unit, extracting outputs from the butterfly engines and passing them to the Streamer, which sends the processed samples back to memory.

The accelerator outputs data in bit-reversed order, requiring special care during the final \ac{FFT} stage to avoid banking conflicts when reordering. As a result, only two bit-reversed samples can be written per cycle, temporarily stalling the pipeline to prevent overwriting meaningful data in the accelerator registers. This limitation allows the full output bandwidth to be utilized only in C64 \ac{FFT}s, while in C32 \ac{FFT}s, only half of the bandwidth can be exploited.

To optimize power consumption, a two-level clock-gating mechanism has been implemented. At the cluster level, a clock-gating cell completely disables the accelerator when not in use, controlled via a dedicated software-configurable register. At the accelerator level, the accelerator remains inactive, except for the control unit responsible for programming, until execution is triggered. Once the job is completed, the accelerator automatically clock-gates its functional units.

%% file: 04-Physical_Implementation_and_measurements.tex
%
\begin{figure}[!t]
    \centering
    \includegraphics[width=1\linewidth]{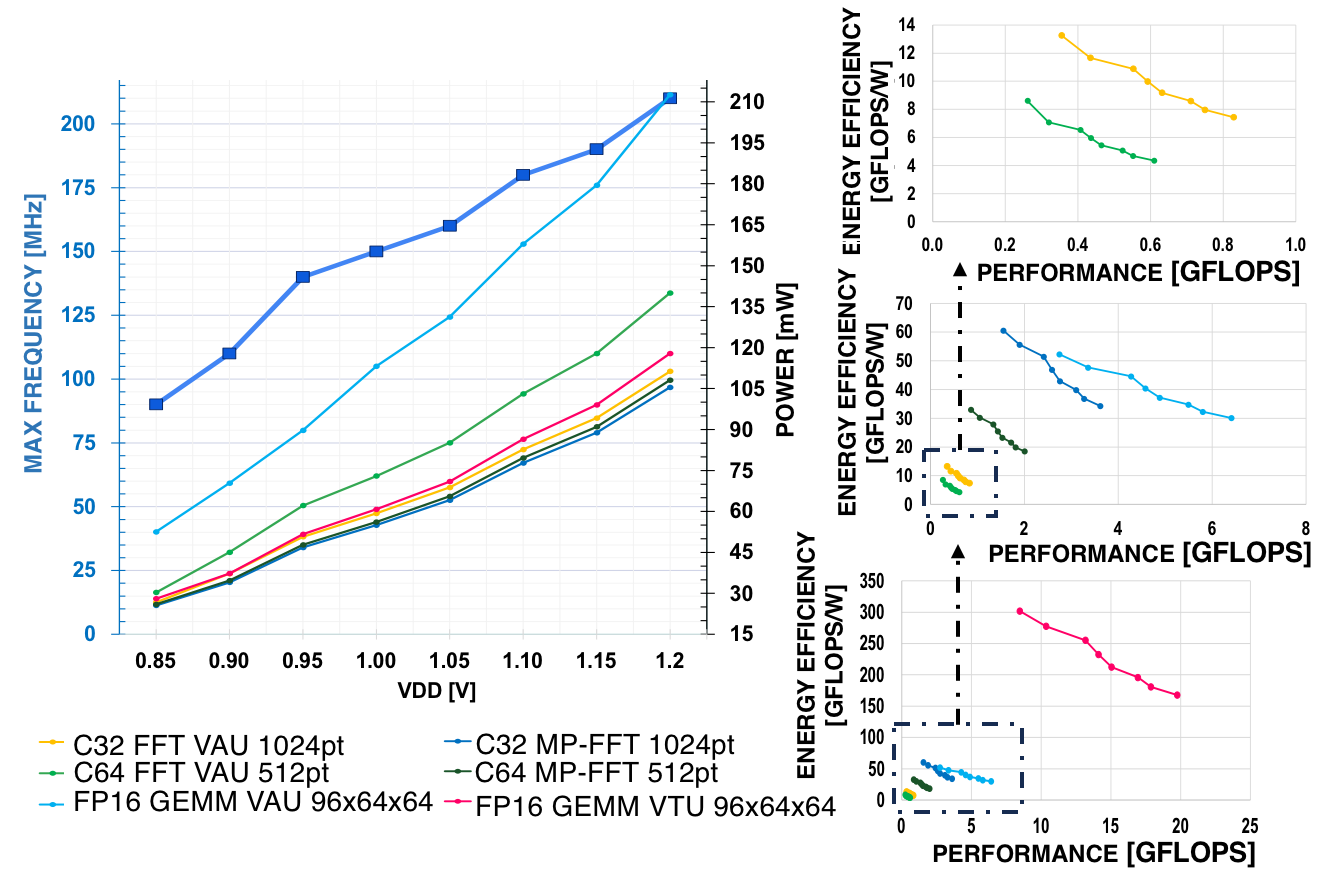}
    \caption{Power - Frequency and  Energy- Performance sweeps for FFT and MatMul Kernels.}
    \label{fig:sweep}
\end{figure}

\begin{figure*}[t!]
    \centering
    \includegraphics[width=1\linewidth]{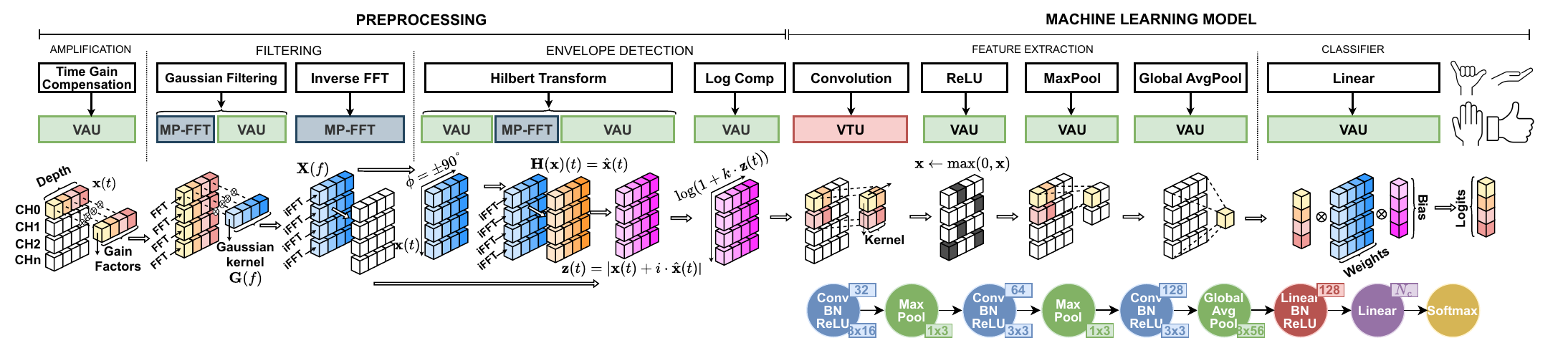}
    \caption{Wearable Ultrasound A-Mode gesture recognition pipeline on Maestro SoC.}
    \label{fig:wus_processing}
\end{figure*}

\section{Physical Implementation and measurements} \label{sec:Physical Implementation}
Fig. \ref{fig:micrograph} shows the chip micrograph and the area breakdown of the Cluster in the Maestro \ac{SoC}, highlighting the main building blocks described in Section \ref{sec:Maestro Architecture}.
The \ac{SoC} is implemented in TSMC 65nm CMOS technology and occupies 12mm$^2$ (3mm $\times$ 4mm) of die area, of which the Cluster domain occupies 7.28mm$^2$ (2.8mm $\times$ 2.6mm).
It has been synthesized with Synopsys Design Compiler 2022.03, while Place \& Route has been performed with Cadence Innovus 21.17, and chip finishing with Calibre 2022.3.
The chip has been tested and characterized using an Advantest \ac{SoC} hp9300 integrated circuit testing device.

Fig. \ref{fig:sweep} illustrates the maximum operating frequency and power consumption across a voltage range of 0.85V to 1.2V for representative computational workloads operating on low data precision FP16-FP32. These workloads include key tasks for embedded machine learning and frequency domain applications, executed on the \ac{VAU}, \ac{VTU}, and the  MP-\ac{FFT} accelerator. The Cluster operates within a frequency spectrum from 90MHz at 0.85V to 210MHz at 1.2V, with range power consumption varying from 25mW to 212mW.
%
%

An additional overview of the kernels' performance and energy efficiency across all operating points is provided in Fig. \ref{fig:sweep}. Fig. \ref{fig:kernel_perf} presents a more detailed analysis, focusing on peak performance and energy efficiency for each kernel configuration across different workload sizes for tensor and \ac{FFT} operations. For MatMul, it evaluates a smaller 12$\times$16 matrix multiplication, while for \ac{FFT}, it examines different data precisions across various point sizes.

The \ac{VTU} achieves  98\% utilization of its internal \ac{FMA}s, delivering near-to-ideal performance of 47 actual FMA/cycle on a 96×64 FP16 MatMul (out of 48 ideal \ac{FMA}/cycle). This represents a 3$\times$ and 5.78$\times$ improvement over the Vector FPUs, achieving, respectively, a peak performance of 19.8GFLOPS and an energy efficiency of 301.7GFLOPS/W.
Even with 50\% utilization on a 12×16 FP16 MatMul, the \ac{VTU} enhances performance by up to 15× and energy efficiency by up to 12× for MatMul kernels with the same precision compared to \ac{VAU}.

Similarly, the \ac{FFT} processor outperforms the \ac{VU} in both performance and efficiency across various \ac{FFT} workloads by up to 4.80$\times$ and 4.68$\times$, respectively on C32 128-point. Notably, the C32 \ac{FFT} 1024-point configuration achieves a peak performance of 3.6GFLOPS and a peak efficiency of 60.6GFLOPS/W.

%% file: 05-End-to-end_Ultrasound_Application.tex
\section{End-to-end Ultrasound Application} \label{sec:End-to-end Ultrasound Application}

In this section, we present the hardware mapping of the \ac{WUS} application pipeline to demonstrate the efficiency of the proposed \ac{SoC}.
The application is inspired by the work by Zeng et al.~\cite{10208224} and consists of a preprocessing chain followed by a \ac{CNN} for gesture classification. Our implementation is designed to process and recognize hand gestures using 8 \ac{US} transducers, each producing 512 echo samples.

\begin{figure*}[t!]
    \centering
    \includegraphics[width=1\linewidth]{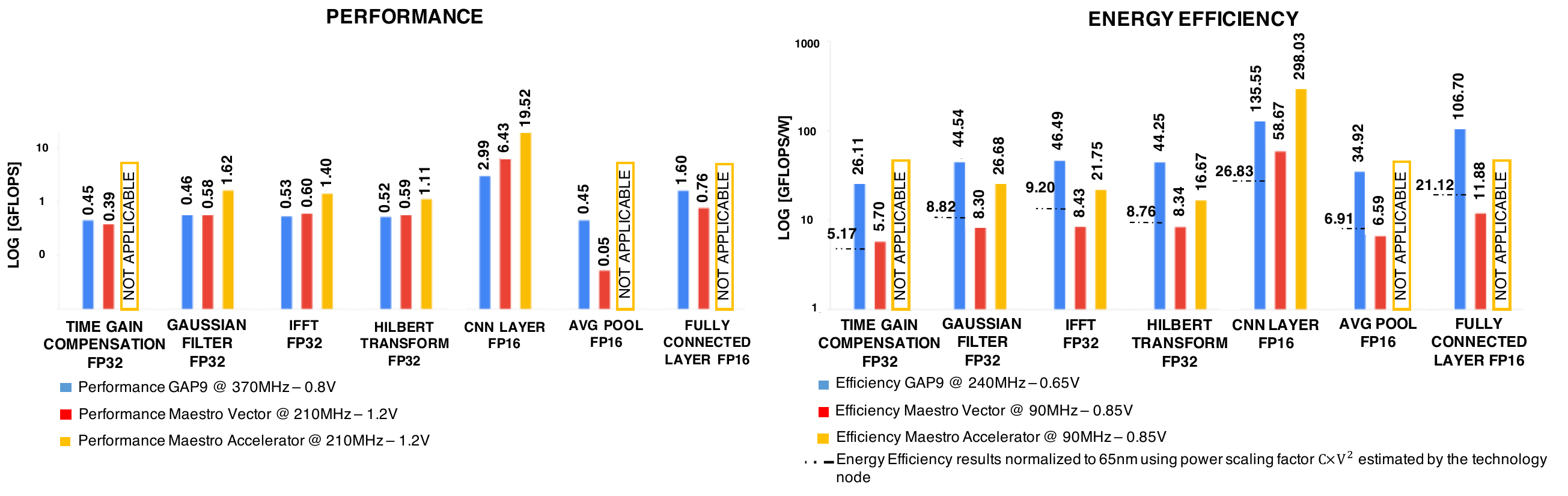}
    \caption{Wearable Ultrasound A-Mode performance and energy efficiency comparison, showing Maestro executing on the VAU (red), VAU + MP-FFT or VTU (yellow), and GAP9 (blue).}
    \label{fig:wus_perf_eff}
\end{figure*}

\subsection{Wearable Ultrasound Preprocessing} \label{subsec: WUS Preprocessing}
The preprocessing in Fig.~\ref{fig:wus_processing} utilizes \ac{DSP} algorithms to perform amplification, filtering, and envelope detection on A-mode \ac{US} signal acquisition. Incoming \ac{US} signals are first amplified using \ac{TGC} and filtered with \ac{GF} to eliminate noise and enhance relevant information. By leveraging vector operations, the \ac{TGC} computation processes 512 data samples simultaneously, significantly improving throughput.
Implementing the Gaussian filter involves applying a Fourier transform of the \ac{US} signal using the \ac{FFT} accelerator and then multiplying transformed data by a Gaussian kernel through the \ac{VU}, leveraging parallel processing. Finally, the filtered signals are reconstructed via an \ac{iFFT}, efficiently executed on the \ac{FFT} accelerator.
Envelope detection leverages the \ac{HT}, with computations distributed between the \ac{VU} and the \ac{FFT} accelerator. The process begins with the \ac{VU} applying a phase shift to the filtered output. Next, the \ac{FFT} accelerator performs \ac{iFFT} to reconstruct the signal. Finally, the \ac{VU} calculates the magnitude to extract the envelope, completing the detection phase.
Lastly, the \ac{VU} performs logarithmic compression on the envelope derived from the \ac{HT} process, reducing the dynamic range and enhancing the visibility of key signal features.

\subsection{Wearable Ultrasound Machine Learning Model} \label{subsec: WUS Machine Learning Model}
The \ac{CNN} model derived from~\cite{10208224} is illustrated in Fig.~\ref{fig:wus_processing}. It comprises three 2D convolutional layers, each followed by \ac{BN}, ReLU activation, and \ac{MP}, progressively compressing data along the echo dimension while increasing the number of feature maps to 128. In particular, convolutional and \ac{BN} layers are fused together and executed on the \ac{VTU}, whereas \ac{MP} are mapped to the \ac{VU}.
Then, global average pooling is applied, resulting in a 128-dimensional vector, which is passed through a \ac{FC} layer---also followed by \ac{BN} and ReLU. Likewise, the \ac{FC} and \ac{BN} layers are fused together and executed on the \ac{VU}.
Finally, an additional linear layer followed by a softmax activation function produces the probability distribution over the gesture categories.

\subsection{Wearable Ultrasound Mapping and Performance Results} \label{subsec: End-to-End Results}

The capabilities of Maestro have been evaluated by running the \ac{WUS} application on the \ac{VAU} alone and, where applicable, leveraging the combined execution of the Vector-Tensor-FFT engines. The evaluation focuses on latency, performance, and energy efficiency, comparing results against GAP9, a commercial version of the VEGA \ac{SoC}~\cite{VEGA}, developed by GreenWaves Technologies and fabricated in GlobalFoundries 22nm FDX technology, with operating ranges from \SI{0.65}{V} @ \SI{240}{MHz} to \SI{0.8}{V} @ \SI{370}{MHz}.

The preprocessing phase is evaluated using FP32 precision on both Maestro and GAP9, analyzing the execution time and performance of each individual preprocessing step for a single \ac{US} channel. This approach enables a precise comparison of Gaussian Filtering, inverse FFT, and Hilbert Transform on both platforms. The results, shown in Fig. \ref{fig:wus_perf_eff}, highlight how Maestro outperforms GAP9 in these accelerated tasks, particularly when utilizing the \ac{FFT} accelerator, which significantly enhances preprocessing efficiency in terms of normalized energy metrics.

The \ac{ML} model evaluation is performed using FP16 precision on the entire gesture recognition pipeline, where the eight \ac{US} channels are preprocessed before feature extraction. The \ac{VTU} demonstrates higher performance in \ac{CNN}-based feature extraction. Despite the technology scaling difference (TSMC 65nm CMOS vs. GlobalFoundries 22nm FDX), which impacts energy efficiency, the \ac{VTU} remains 2$\times$ more efficient in \ac{CNN} tasks, achieving 298.03GFLOPS/W.

\begin{figure}[t!]
    \centering
    \includegraphics[width=1\linewidth]{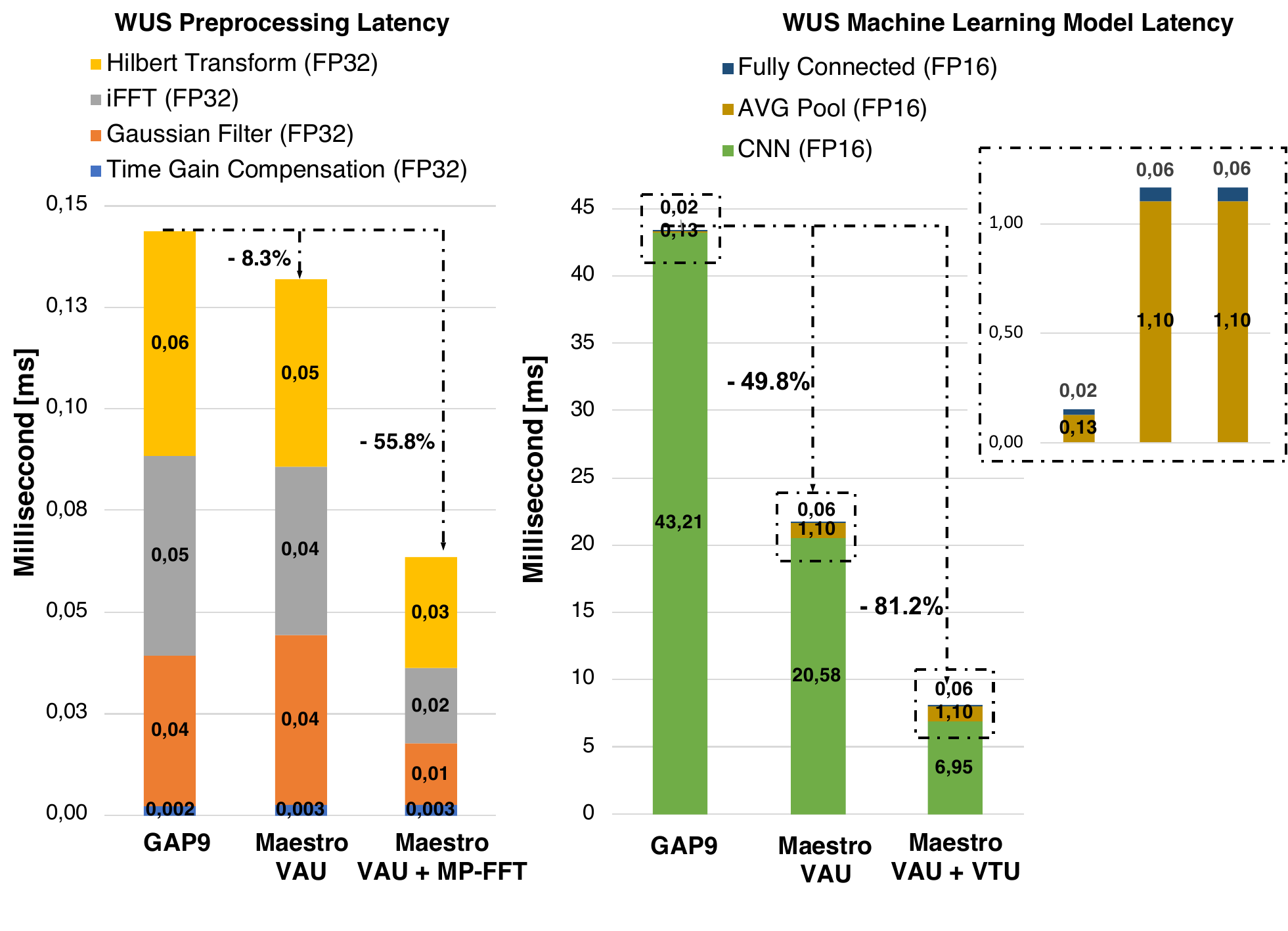}
    \caption{Wearable Ultrasound A-Mode execution latency on GAP9, Maestro VAU only and Maestro VAU + MP-FFT or VTU.}
    \label{fig:wus_latency}
\end{figure}

Fig. \ref{fig:wus_latency} presents the execution latency of the \ac{WUS} application. While Maestro runs at 1.76$\times$ slower frequency than GAP9, its \ac{VU} and the high-performance \ac{FFT} accelerator enable the preprocessing phase to be completed 2.4$\times$ faster than GAP9, demonstrating the efficiency of the architecture in handling frequency-domain and signal processing tasks.

The majority of the ML execution time is consumed by \ac{CNN} computations, whereas average pooling and fully connected layers account for less than 1\% of the total execution time considering GAP9 execution. The primary reason for the suboptimal performance of the average pooling layer on the \ac{VU} is its reliance on reduction operations, which are not well-suited to leverage the \ac{VU}'s capabilities. Additionally, the vector-matrix multiplication used in fully connected layers reduces the parallelism of the \ac{VU}, limiting its ability to achieve maximum throughput. However, since these two layers contribute only 1\% to the total computation time, the overall performance impact remains minimal. The computational power of the \ac{VTU} effectively compensates for the performance limitations of the average pooling and fully connected layers.  While the \ac{VU} achieves a 2$\times$ speedup compared to GAP9, the \ac{VTU} accelerates \ac{CNN}-based feature extraction by up to 6.2$\times$, reducing the \ac{ML} execution time by 81\%.
Overall, the entire \ac{WUS} application is executed 5$\times$ faster on the highly specialized architecture of Maestro.

\subsection{Wearable Ultrasound Maestro Platform: Power and Lifetime Analysis} \label{subsec: Wearable Ultrasound Maestro Platform: Power and Lifetime Analysis}

\begin{figure}[t!]
    \centering
    \includegraphics[width=1\linewidth]{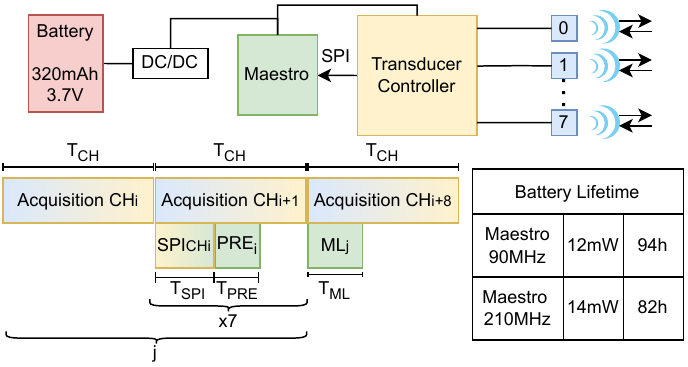}
    \caption{Power and Lifetime Analysis of Maestro-based \ac{WUS} processing platform.}
    \label{fig:wus_maestro_system}
\end{figure}

We conceptualized the Maestro \ac{WUS} platform, inspired by Frey et al. \cite{wulpus}, to enable efficient on-device \ac{US} processing. The analytical study in Fig. \ref{fig:wus_maestro_system} considers a 320mAh, 3.7V LiPo battery powering a system composed of eight \ac{US} transducers, a transducer controller for managing data acquisition, and the Maestro \ac{SoC}, which receives input samples via SPI. The platform is configured to acquire 512 echo samples per frame at a frame rate of 39Hz.

By pipelining the acquisition phase with the preprocessing and \ac{ML} model, described in Section \ref{subsec: WUS Preprocessing} and Section \ref{subsec: WUS Machine Learning Model}, respectively, this system fully supports energy-efficient \ac{WUS} edge processing. The platform achieves up to 94 hours of battery life at Maestro's highest energy efficiency point, consuming 12mW and 2.5mJ, while at peak performance conditions, it operates for 82 hours, consuming 14mW and 2.9mJ.

\input{06.1-SoA_Comparison}

%% file: 06.1-SoA_Comparison.tex

\begin{table*}[t!]
\centering
\begin{threeparttable}
\setlength{\tabcolsep}{2pt} 
\caption{Comparison with SoA SoCs.}
\label{tab:comp_soa}
\begin{tabular}{|c|c|c|c|c|c|c|}
    \hline
    &
        \begin{tabular}[c]{@{}c@{}}\cite{YUN}\\\textit{YUN}\end{tabular} &
        \begin{tabular}[c]{@{}c@{}}\cite{Echoes}\\\textit{ECHOES}\end{tabular} &
        \begin{tabular}[c]{@{}c@{}}\cite{8240277}\\\textit{LTE/Wi-Fi FFT SoC}\end{tabular} &
        \begin{tabular}[c]{@{}c@{}}\cite{Garofalo2023DARKSIDEAH}\\\textit{DARKSIDE}\end{tabular} &
        \begin{tabular}[c]{@{}c@{}}\cite{VEGA}\\\textit{VEGA}\end{tabular} &
        \textit{\begin{tabular}[c]{@{}c@{}}MAESTRO\\(this work)\end{tabular}} \\ \hline
    \textit{Technology} & \mycomment{Intel 16-nm FinFet} 65-nm TSMC & 65-nm TSMC & 16-nm FinFET & 65-nm TSMC & 22-nm CMOS FDSOI & 65-nm TSMC \\ \hline
    \begin{tabular}[c]{@{}c@{}}\textit{Die size}\\$[\text{mm}^2]$\end{tabular} & \mycomment{24.01 &} 6 & 4 & 1.28 & 12 & 12 & 12 \\ \hline
    \textit{Application} &
        Vector &
        IoT+FFT &
        IoT+FFT &
        \begin{tabular}[c]{@{}c@{}}IoT+NSAA+\\DNN+QNNs\end{tabular} &
        \begin{tabular}[c]{@{}c@{}}IoT+NSAA+DNN\end{tabular} &
        \begin{tabular}[c]{@{}c@{}}IoT+Radar+FFT\end{tabular} \\ \hline
    \textit{Cores} &
        RV64GCV0.9 &
        \begin{tabular}[c]{@{}c@{}}1xRV32FXpulp+FFT\end{tabular} &
        Rocket-Chip &
        \begin{tabular}[c]{@{}c@{}}8xRISCY-NN\\RVC32IMFXpulpNN2\end{tabular} &
        \begin{tabular}[c]{@{}c@{}}10xRI5CY\\RVC32IMF-Xpulp\end{tabular} &
        \begin{tabular}[c]{@{}c@{}}1xRV32IMC\\+ 1xRVV 1.0 ZVE64d\end{tabular} \\ \hline
    \begin{tabular}[c]{@{}c@{}}\textit{Supply Voltage}\\$[\text{V}]$\end{tabular} & \mycomment{0.55--1 &} 0.85--1.2 & 0.9--1.2 & 0.57--0.9 & 0.75--1.2 & 0.5--0.8 & 0.85--1.2 \\ \hline
    \begin{tabular}[c]{@{}c@{}}\textit{Frequency}\\$[\text{MHz}]$\end{tabular} &
        100--280 &
        150--350 &
        40--320 &
        40--290 &
        32e-3--450 &
        90--210 \\ \hline
    \begin{tabular}[c]{@{}c@{}}\textit{Power range}\\$[\text{mW}]$\end{tabular} &
        57--330 &
        8--133.5 &
        0.46--22 &
        12--213 &
        \textbf{1.7e-3--49.4} &
        21--218 \\ \hline
    \textit{FP precision} &
        FP64, FP32 &
        FP32 &
        FP16, FP32, FP64&
        FP16, FP32 &
        bfloat16, FP16, FP32 &
        \begin{tabular}[c]{@{}c@{}}\textbf{FP64, FP32, FP16, BF16}\\ \textbf{FP8 (4,3), FP8 (5,2)}\end{tabular} \\ \hline
    \textit{FPUs} &
        4 $\times$ FP64 &
        1 $\times$ FP32 &
        1 $\times$ FP64 &
        \begin{tabular}[c]{@{}c@{}}4 $\times$ FP32 +\\ 32 $\times$ FP16\end{tabular} &
        bfloat16, 4 $\times$ FP32 &
        \begin{tabular}[c]{@{}c@{}}\textbf{4 $\times$ FP64 +}\\ \textbf{48 $\times$ FP16}\end{tabular} \\ \hline
    \textit{\begin{tabular}[c]{@{}c@{}}FFT Accelerator\\Peak Perf\end{tabular}} &
        - &
        \begin{tabular}[c]{@{}c@{}}10.16 GOPS\tnote{A,4}\\ 5.8 GOPS\tnote{B,3}\\3.2 GOPS\tnote{D,1}\end{tabular} &
        8.60 GOPS\tnote{C,2} &
        - &
        - &
        \textbf{\begin{tabular}[c]{@{}c@{}}3.6 GFLOPS\tnote{B,3}\\2 GFLOPS\tnote{D,1}\end{tabular}} \\ \hline
    \textit{\begin{tabular}[c]{@{}c@{}}FFT Accelerator\\Peak Efficiency \end{tabular}} &
        - &
        \begin{tabular}[c]{@{}c@{}}200 GOPS/W\tnote{A,4}\\89.5 GOPS/W\tnote{B,3}\\41.6 GOPS/W\tnote{D,1}\end{tabular} &
        380.55 GOPS/W\tnote{C,2} &
        - &
        - &
        \textbf{\begin{tabular}[c]{@{}c@{}}60.6 GFLOPS/W\tnote{B,3}\\33 GFLOPS/W\tnote{D,1}\end{tabular}} \\ \hline
    \begin{tabular}[c]{@{}c@{}}\textit{Tensor Peak Perf}\\$[\text{GFLOPS}]$\end{tabular} &
        - &
        - &
        - &
        18.2\tnote{b} &
        - &
        \textbf{19.8\tnote{b}} \\ \hline
    \begin{tabular}[c]{@{}c@{}}\textit{Tensor Peak Eff}\\$[\text{GFLOPS/W}]$\end{tabular} &
        - &
        - &
        - &
        300\tnote{b} &
        - &
        \textbf{301.7\tnote{b}} \\ \hline
    \begin{tabular}[c]{@{}c@{}}\textit{FP Peak Perf}\\$[\text{GFLOPS}]$\end{tabular} &
        \begin{tabular}[c]{@{}c@{}}2.83\tnote{d}\\\textbf{5.70\tnote{c}}\end{tabular} &
        0.16\tnote{c} &
        320 MFLOPS\tnote{*,b} &
        \begin{tabular}[c]{@{}c@{}}1.02\tnote{b}\\1.02\tnote{c}\end{tabular} &
        \begin{tabular}[c]{@{}c@{}}3.3\tnote{b}\\2\tnote{c}\end{tabular} &
        \begin{tabular}[c]{@{}c@{}}12.6\tnote{a}\\\textbf{6.6\tnote{b}}\\3.28\tnote{c}\end{tabular} \\ \hline
    \begin{tabular}[c]{@{}c@{}}\textit{FP Peak Eff}\\$[\text{GFLOPS/W}]$\end{tabular} &
        \begin{tabular}[c]{@{}c@{}}10.8\tnote{d}\\23.4\tnote{c}\end{tabular} &
        9.68\tnote{c} &
        - &
        \begin{tabular}[c]{@{}c@{}}6.2\tnote{b}\\5\tnote{c}\end{tabular} &
        \begin{tabular}[c]{@{}c@{}}\textbf{129\tnote{b}}\\\textbf{79\tnote{c}}\end{tabular} &
        \begin{tabular}[c]{@{}c@{}}\textbf{114.10\tnote{a}}\\60.23\tnote{b}\\29.85\tnote{c}\end{tabular} \\ \hline
\end{tabular}
\begin{tablenotes}
    \item 
        \textbf{FP Precision:}
        a) FP8 -
        b) FP16 -
        c) FP32 -
        d) FP64
        \textbf{FFT format:}
        A) C16 -
        B) C32 -
        C) C48 -
        D) C64
    
\end{tablenotes}
\begin{tablenotes}
    \item
        \textbf{FFT points:}
        1) 512 -
        2) 972 -
        3) 1024 -
        4) 2048
        \textbf{Notes:}
        (*) Estimated from the paper 
\end{tablenotes}

\end{threeparttable}
\end{table*}

%% file: 06-SoA_Comparison.tex
\section{SoA Comparison} \label{sec:SoA Comparison}
Table \ref{tab:comp_soa} compares the Cluster of Maestro against a selection of state-of-the-art \ac{SoC} with vector, tensor, and \ac{FFT} processors. 
Compared to YUN \cite{YUN}, Maestro meets the requirement of low power consumption (21mW - 212mW) for a wearable device, while YUN relies on a less efficient architecture tuned for \ac{HPC}, resulting in lower efficiency (-28\% compared to FP32 execution in Maestro’s vector unit and 2.57$\times$ compared to FP16). The \ac{FFT} accelerators in \cite{Echoes,8240277} are limited to fixed-point precision, making them less suitable for \ac{WUS} applications with wide data ranges. Maestro’s \ac{FFT} accelerator supports floating-point data, with only ~50\% lower efficiency than ECHOES \cite{Echoes} in C32int and ~6.3$\times$ lower than \cite{8240277}, which targets a more aggressive technology node (16nm). Maestro’s tensor unit achieves 301.7GFLOPS/W, 19.8GFLOPS@FP16 outperforming YUN \cite{YUN} in terms of peak performance and \cite{VEGA,YUN} in energy efficiency on \ac{GEMM} kernels. Compared to Darkside \cite{Garofalo2023DARKSIDEAH}, Maestro provides similar energy efficiency and slightly better performance with significantly improved flexibility due to the tight Vector-Tensor integration and \ac{FFT} processor tuned for frequency domain processing. Maestro achieves a notable 2.33$\times$ improvement in energy efficiency over VEGA\cite{VEGA} for FP16 tensor operations. While VEGA provides balanced performance across FP16 and FP32, Maestro’s architecture enhances FP16 performance by up to 6$\times$ on the tensor unit and 2$\times$ on the vector unit, with a 0.9$\times$ factor in FP32. Furthermore, in an end-to-end \ac{WUS} application, Maestro's \ac{VAU} achieves a 2$\times$ speedup. In contrast, its \ac{VTU} accelerates \ac{CNN}-based feature extraction by up to 6.2$\times$ compared to VEGA, resulting in an overall 5$\times$ faster \ac{WUS} execution on Maestro. Thus, Maestro paves the way for entirely shifting \ac{US} processing to the edge, reducing reliance on external computation while enhancing latency, power efficiency, and system autonomy for \ac{WUS} applications.

%% file: 07-Conclusion.tex
\section{Conclusion} \label{sec:Conclusion}
We presented Maestro, a specialized \ac{SoC} designed to bring \ac{WUS} processing to the edge, eliminating the need to offload computation to remote PCs. By integrating frequency-domain and tensor acceleration within a compact and efficient embedded vector architecture, Maestro minimizes latency, power consumption, and privacy concerns, making it well-suited for wearable applications.
Fabricated in low-cost TSMC 65nm CMOS technology, Maestro features an RV32-based Cluster featuring compact Vector-Tensor accelerators and a versatile multi-precision \ac{FFT} engine enabling efficient execution of embedded workloads across diverse computational domains within a 212mW power envelope.
The \ac{WUS} application evaluation demonstrates that this \ac{WUS}-optimized architecture significantly accelerates processing while maintaining ultra-low power consumption of 12mW, with an energy consumption of 2.5mJ. Maestro balances general-purpose vector acceleration (60.2 GFLOPS/W, 6.6 GFLOPS @ FP16) with the efficiency of the Tensor Unit (301.7 GFLOPS/W, 19.8 GFLOPS @ FP16) and the \ac{FFT} accelerator (60.6 GFLOPS/W, 3.6 GFLOPS @ C32).


%% file: Authors/Authors_bio_picture.tex
\begin{IEEEbiography}[{\includegraphics[width=1in,height=1.25in,clip,keepaspectratio]{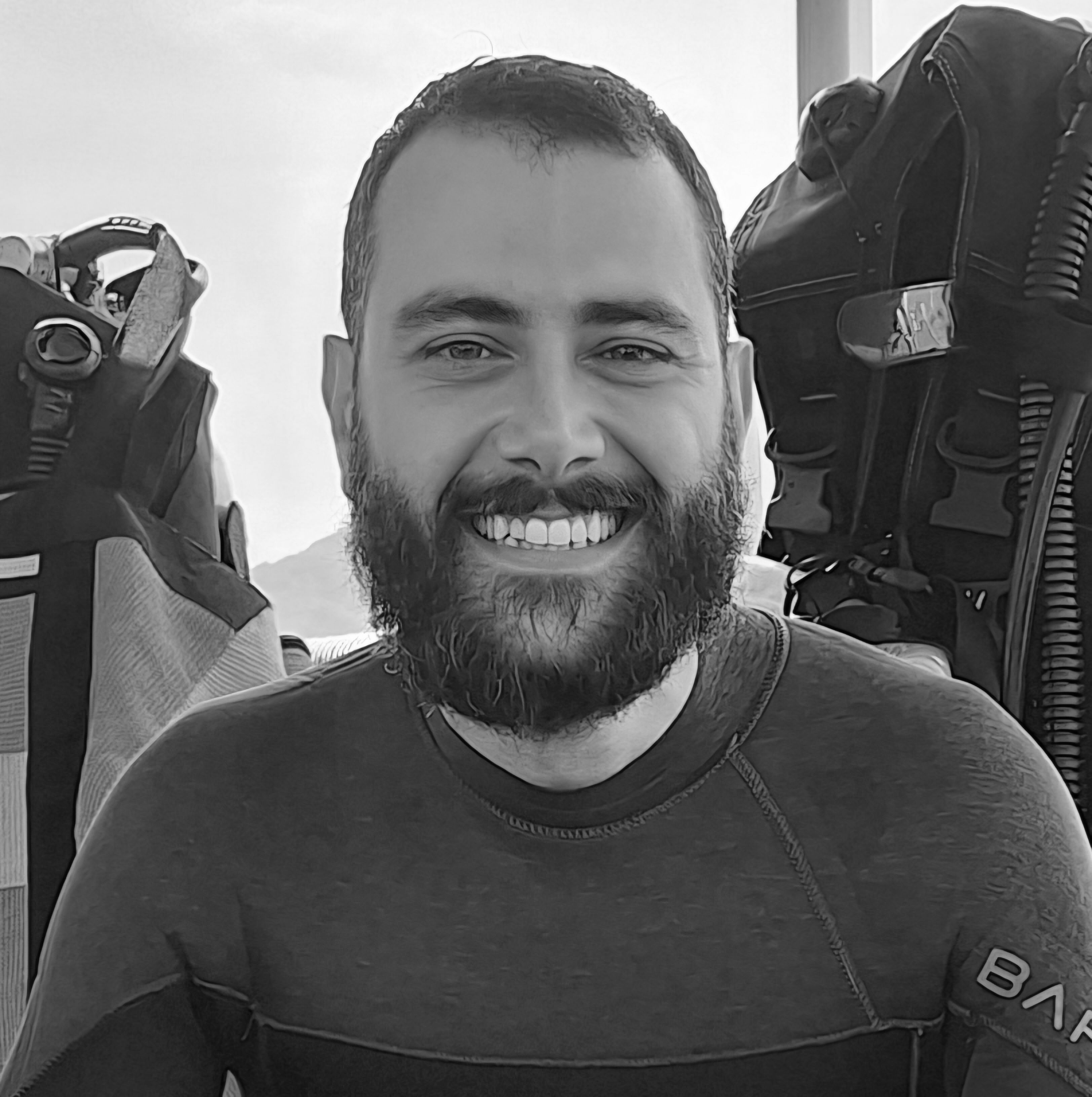}}]{Mattia Sinigaglia} received his Master’s Degree in Electronic Engineering from the University of Bologna in July 2021. He is currently pursuing a Ph. D. in Digital Systems Design at the Department of Electrical and Information Engineering (DEI) of the University of Bologna.  His research interest is semiconductor engineering for ultra-low-power devices, mainly targeting hardware-software co-design of RISC-V-based heterogeneous System-on-Chip.
\end{IEEEbiography}

\vspace{-10mm}

\begin{IEEEbiography}[{\includegraphics[width=1in,height=1.25in,clip,keepaspectratio]{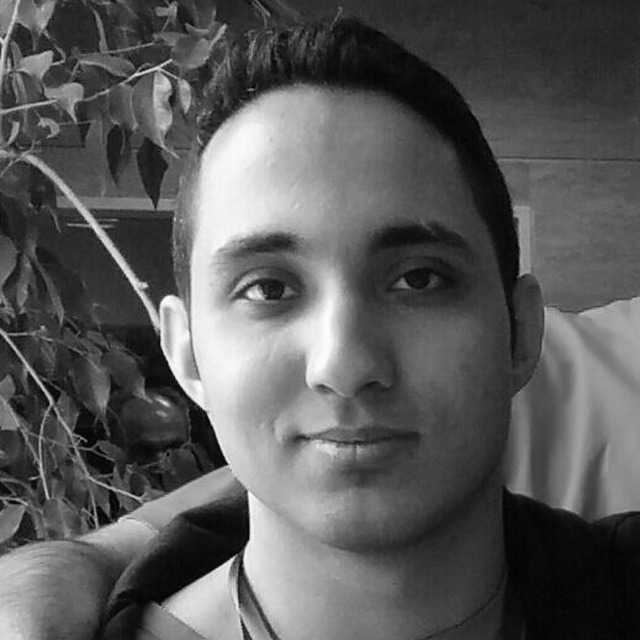}}]{Amirhossein Kiamarzi} received his Master’s degree in Computer Engineering from the University of Shiraz in 2020. He is currently pursuing a Ph.D. in Digital Systems Design at the Department of Electrical and Information Engineering (DEI), University of Bologna. His research focuses on modeling RISC-V-based hardware cores and accelerators.
\end{IEEEbiography}

\vspace{-10mm}

\begin{IEEEbiography}[{\includegraphics[width=1in,height=1.25in,keepaspectratio,clip]{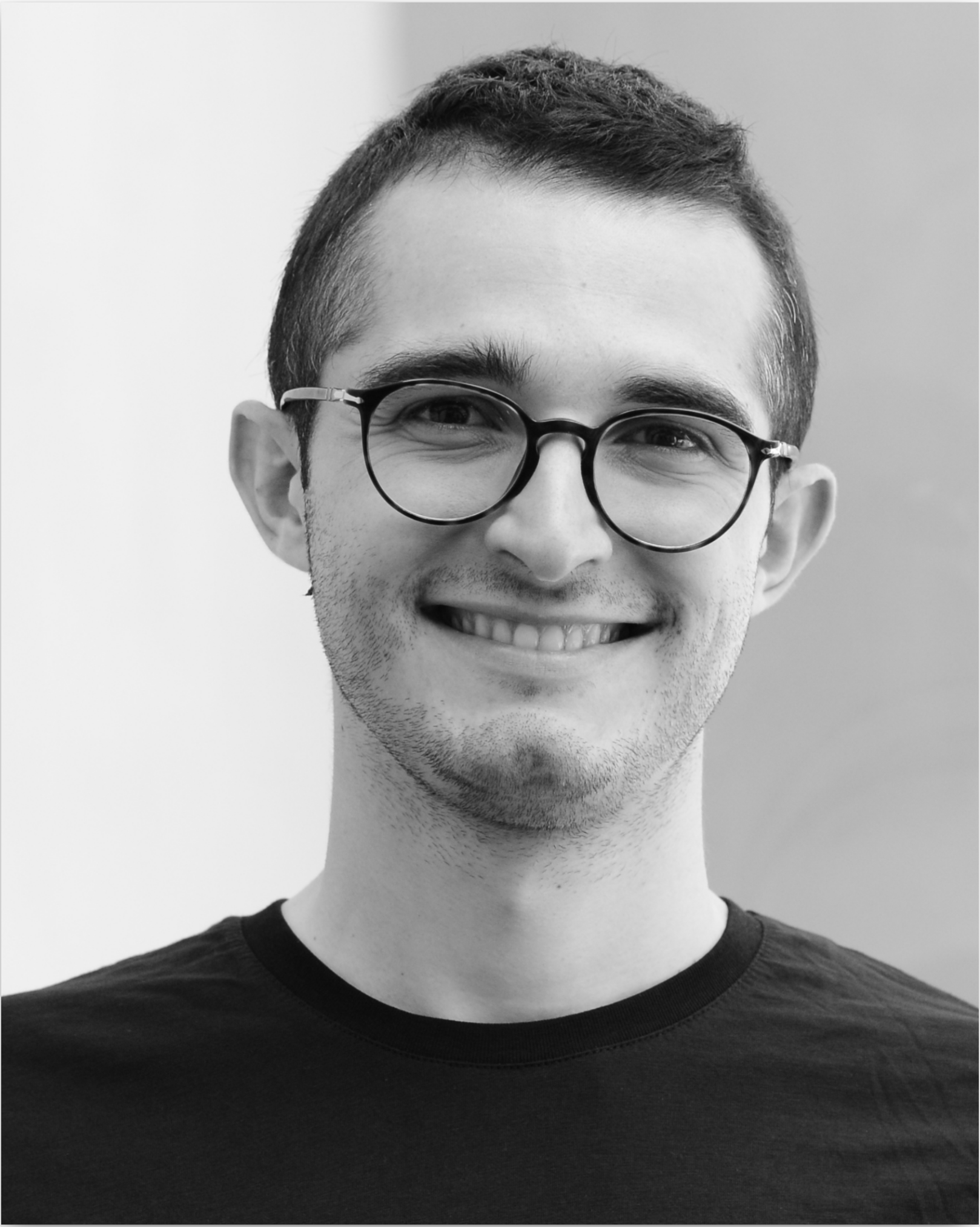}}]{Marco Bertuletti}
received the B.Sc. and M.Sc. degree in Electrical Engineering in Politecnico di Milano, Milano, Italy. He is currently pursuing his Ph.D. at ETH, Zurich, Switzerland, in the Integrated Systems Laboratory (IIS). His main interests are in the design of multi and many-core clusters of RISC-V processors for next-generation telecommunications.
\end{IEEEbiography}

\vspace{-10mm}

\begin{IEEEbiography}[{\includegraphics[width=1in,height=1.25in,clip,keepaspectratio]{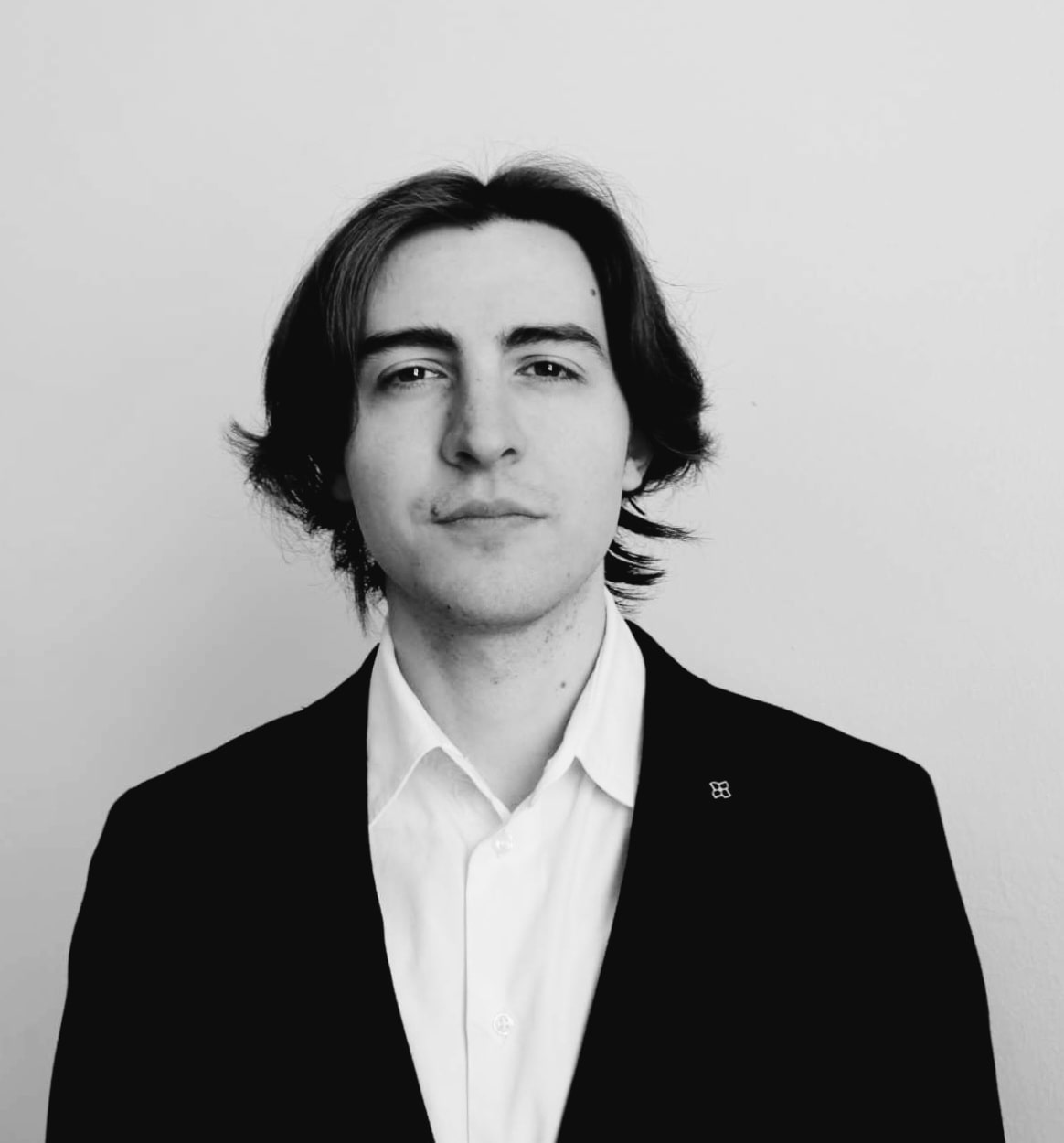}}]{Luigi Ghionda} received his Master’s Degree in Electronic Engineering in March 2024 from the University of Bologna. He is currently a reaserch fellow in the group of Professor Luca Benini at the Department of Electrical and Information Engineering (DEI) of the University of Bologna. His research interests include the design of PULP (Parallel Ultra-Low Power)-based hardware accelerators and the design of RISC-V-based computer architectures for space SoC. 
\end{IEEEbiography}

\vspace{-10mm}

\begin{IEEEbiography}[{\includegraphics[width=1in,height=1.25in,clip,keepaspectratio]{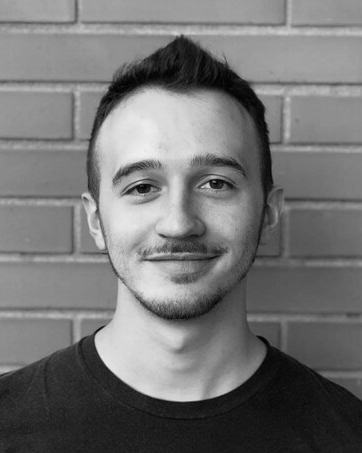}}]{Mattia Orlandi} received his M.Sc. degree in Artificial Intelligence from the University of Bologna, Italy, in 2022. He is currently working toward his Ph.D. in Data Science and Computation under the supervision of Prof. S. Benatti at the Energy-Efficient Embedded Systems Laboratory (EEES Lab), DEI Department, University of Bologna. His research activities involve bio-signal processing with machine learning on low-power computing platforms. He is investigating how to decode EMG signals into spike trains to develop advanced human-machine interfaces.
\end{IEEEbiography}

\vspace{-10mm}

\begin{IEEEbiography}[{\includegraphics[width=1in,height=1.25in,clip,keepaspectratio]{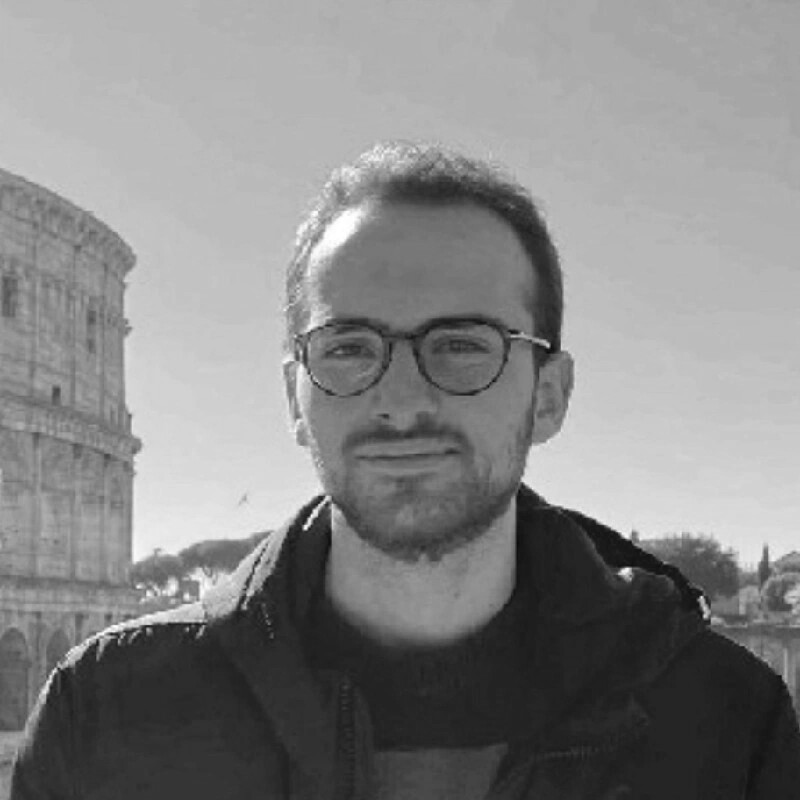}}]{Riccardo Tedeschi} earned his Master’s Degree in Electronic Engineering from the University of Bologna in 2023. He is currently a Ph.D. candidate in Digital Systems Design at the Department of Electrical and Information Engineering (DEI) at the University of Bologna. His research focuses on RISC-V architectures for high-performance embedded platforms, with an emphasis on performance and reliability.
\end{IEEEbiography}

\vspace{-10mm}

\begin{IEEEbiography}[{\includegraphics[width=1in,height=1.25in,clip,keepaspectratio]{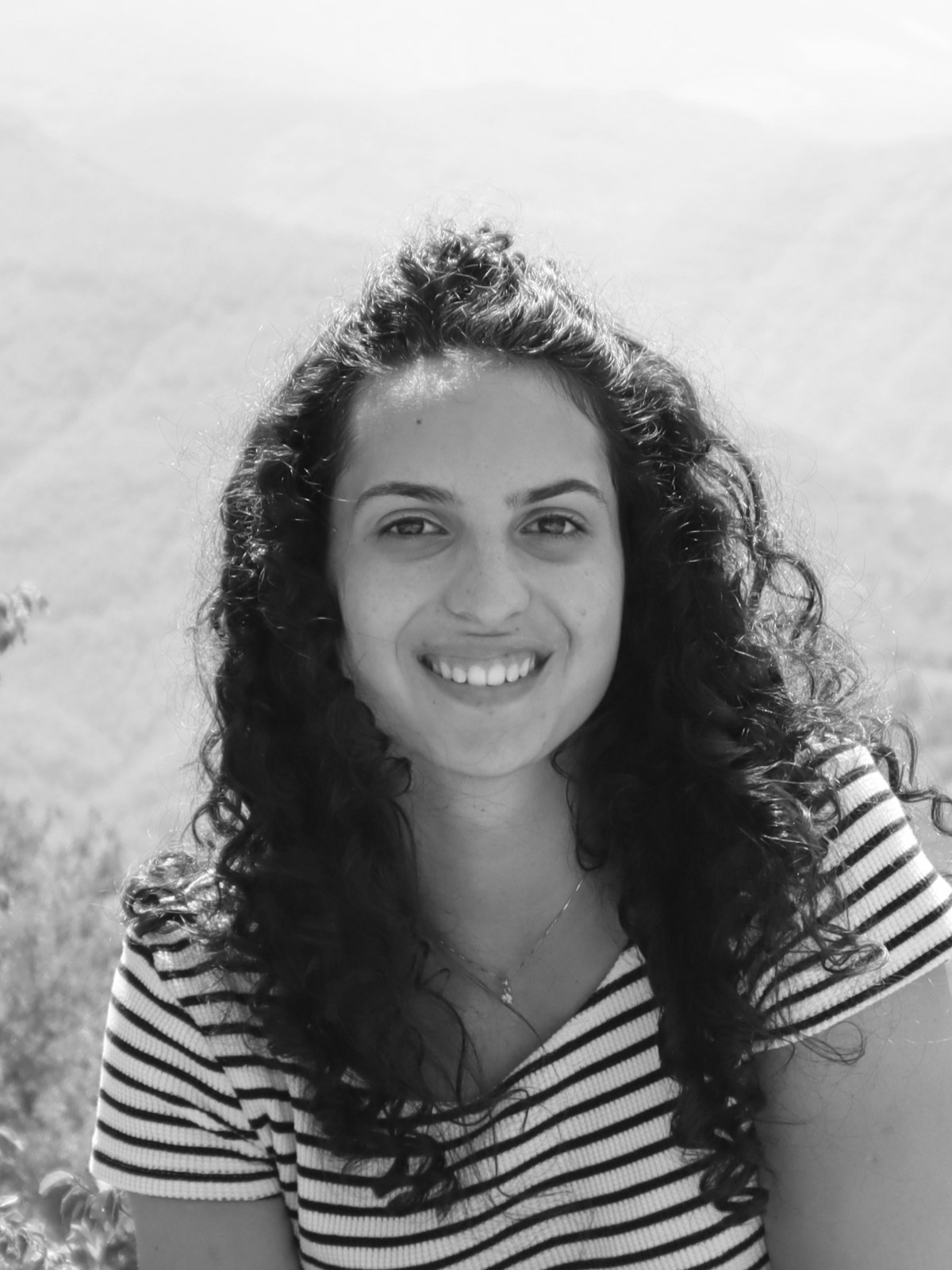}}]{Aurora Di Giampietro} received her Master’s Degree in Electronic Engineering in February 2024 from the University of Bologna. She is currently working as a Research and Development Engineer at Onsemi in Switzerland. Her research interests focus on designing and developing low-power ASICs, with a particular emphasis on energy-efficient solutions for advanced electronic systems.
\end{IEEEbiography}

\vspace{-10mm}

\begin{IEEEbiography}[{\includegraphics[width=1in,height=1.25in,clip,keepaspectratio]{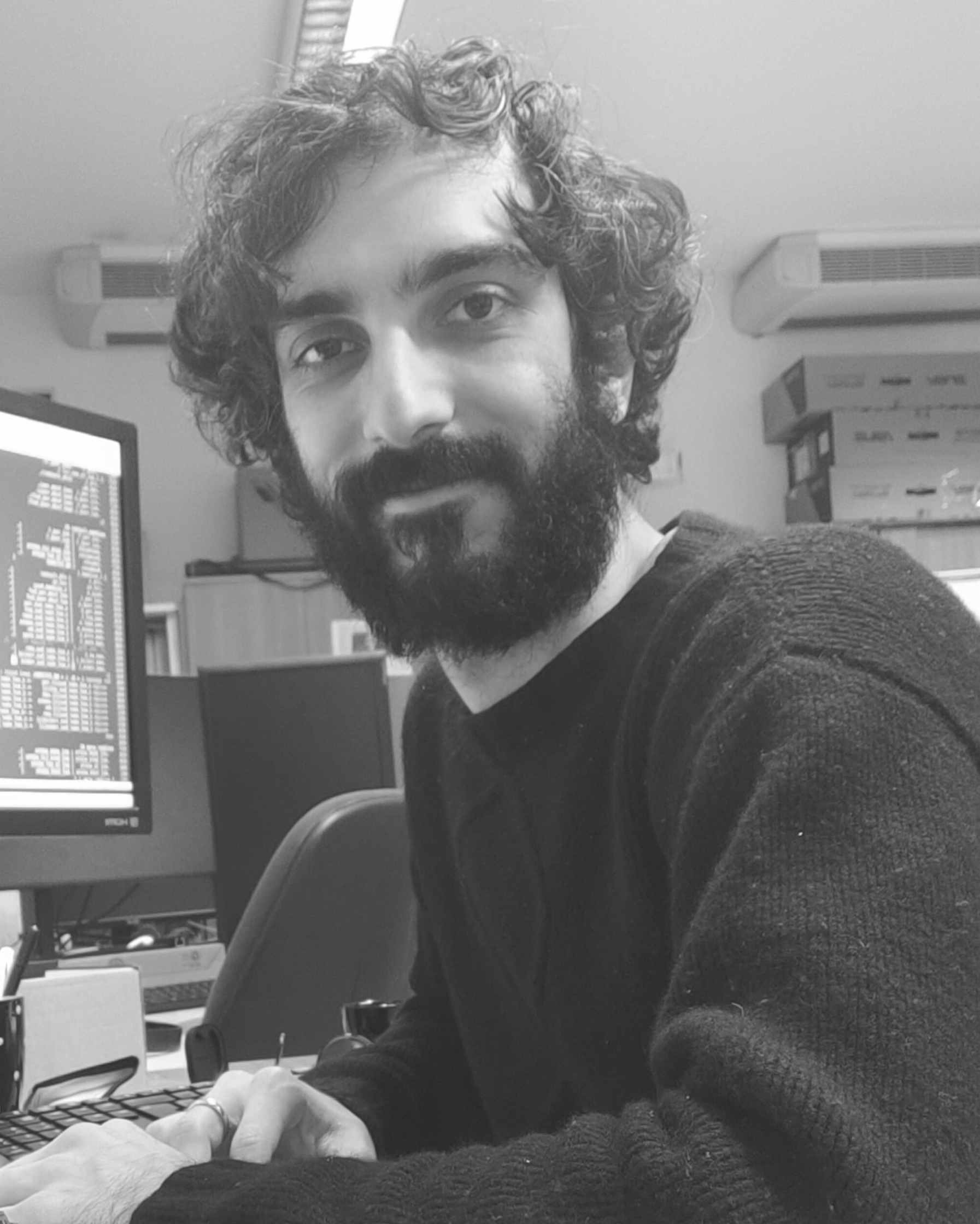}}]{Yvan Tortorella} received his Master’s Degree in Electronic Engineering in October 2021 from the University of Bologna. He is currently pursuing a Ph. D. in Digital Systems Design in the group of Professor Luca Benini at the Department of Electrical and Information Engineering (DEI) of the University of Bologna. His research interests include the design of PULP (Parallel Ultra-Low Power)-based hardware accelerators for ultra-low power Machine Learning and the design of RISC-V-based computer architectures for satellite applications. 
\end{IEEEbiography}

\vspace{-10mm}

\begin{IEEEbiography}[{\includegraphics[width=1in,height=1.25in,clip,keepaspectratio]{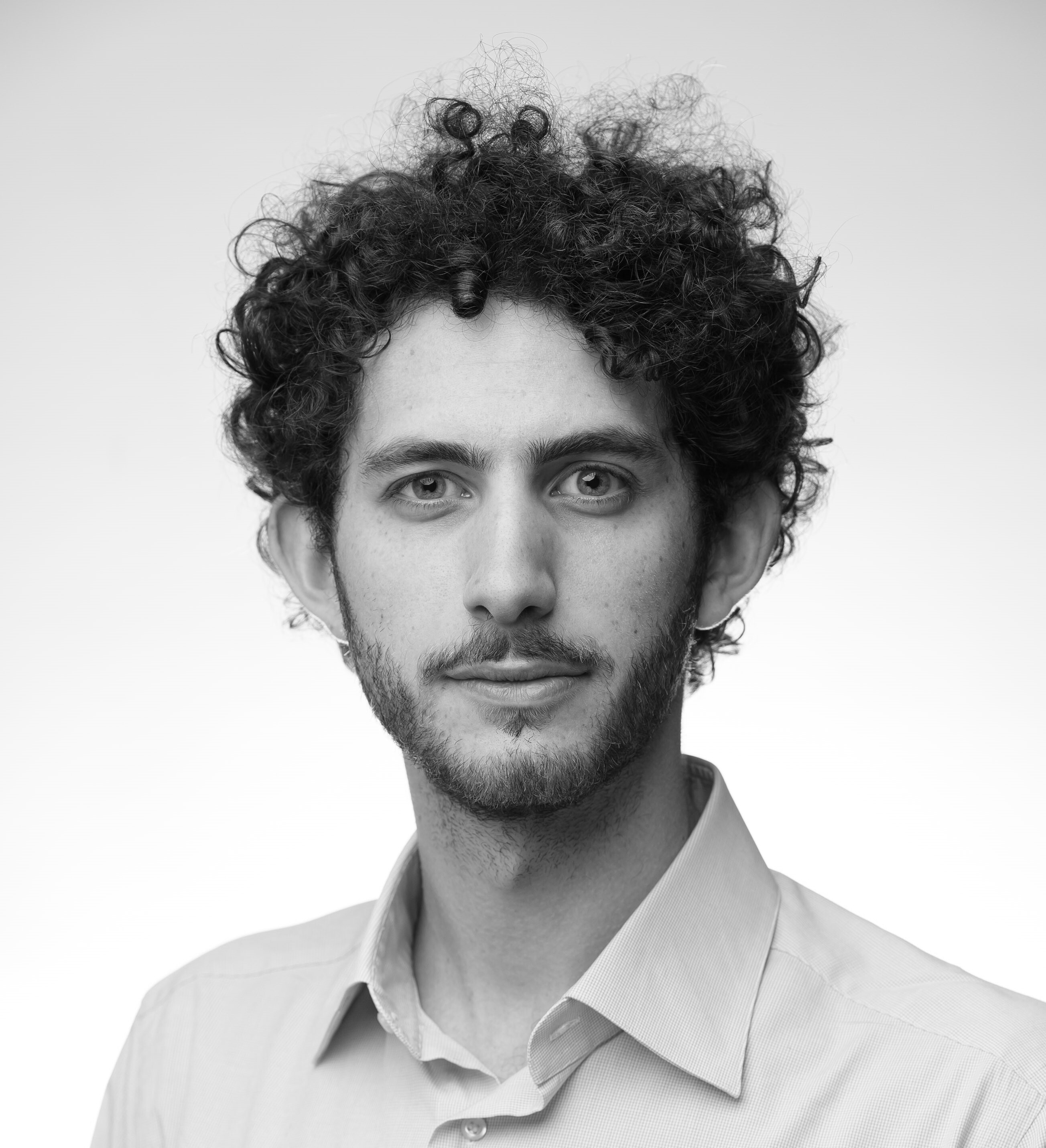}}]{Luca Bertaccini} 
received his Ph.D. degree in Electrical Engineering from ETH Zurich in 2025, where he is currently a postdoctoral researcher in the Digital Circuits and Systems Group of Prof. Benini. His research interests include energy-efficient hardware accelerators, computer arithmetic, and heterogeneous systems-on-chip.
\end{IEEEbiography}

\vspace{-10mm}

\begin{IEEEbiography}[{\includegraphics[width=1in,height=1.25in,clip,keepaspectratio]{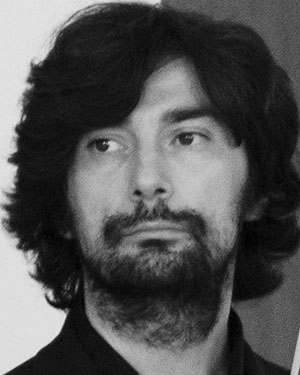}}]{Simone Benatti} received the Ph.D. degree in electrical engineering and computer science from the University of Bologna, Bologna, Italy, in 2016. He has collaborated with several international research institutes and companies. Previously, he worked for 8 years as an Electronic Designer and R\&D Engineer of electromedical devices. He has authored or coauthored more than 50 papers in international peer-reviewed conferences and journals. His research interests focus on energy efﬁcient embedded wearable systems, signal processing, sensor fusion, and actuation systems. 
\end{IEEEbiography}

\vspace{-10mm}

\begin{IEEEbiography}[{\includegraphics[width=1in,height=1.25in,clip,keepaspectratio]{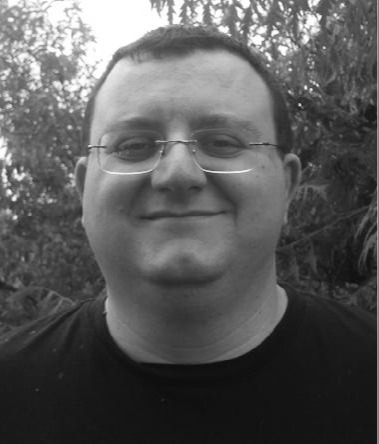}}]{Giuseppe Tagliavini} is a Tenure-Track Assistant Professor at the University of Bologna. His research interests are focused on programming models, orchestration tools, and compiler optimizations for AI-enabled resource-constrained computing platforms, with a strong emphasis on parallel architectures and accelerators in the context of ultra-low-power IoT end nodes. He is a member of the IEEE and ACM societies.
\end{IEEEbiography}

\vspace{-10mm}

\begin{IEEEbiography}[{\includegraphics[width=1in,height=1.25in,clip,keepaspectratio]{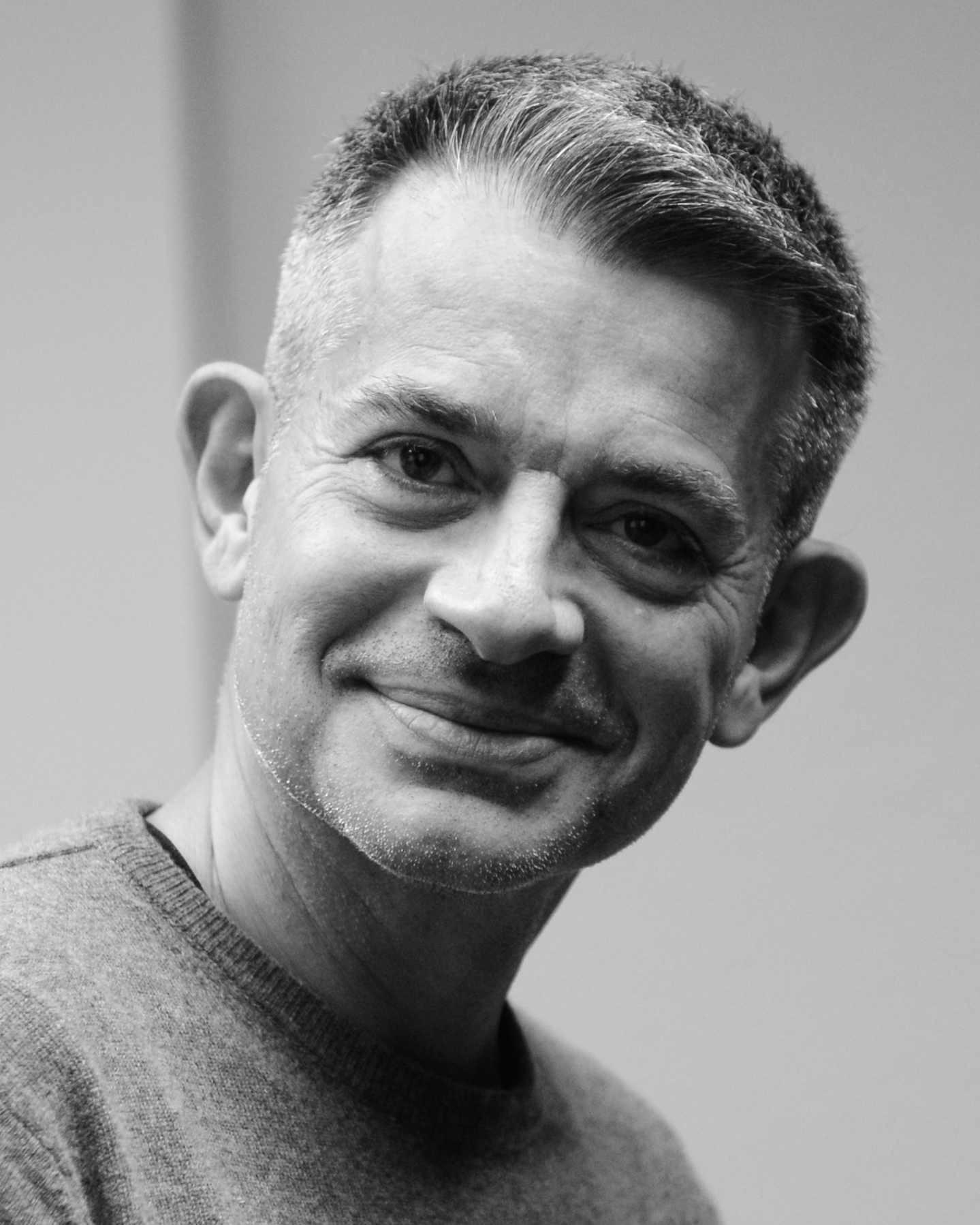}}]{Luca Benini} holds the chair of digital Circuits and systems at ETHZ and is Full Professor at the Universit\`{a} di Bologna. He received a PhD from Stanford University. Dr. Benini’s research interests are in energy-efficient parallel computing systems, smart sensing micro-systems and machine learning hardware. He has published more than 1000 peer-reviewed papers and five books. He is a Fellow of the ACM and a member of the Academia Europaea.
\end{IEEEbiography}

\vspace{-10mm}

\begin{IEEEbiography}[{\includegraphics[width=1in,height=1.25in,clip,keepaspectratio]{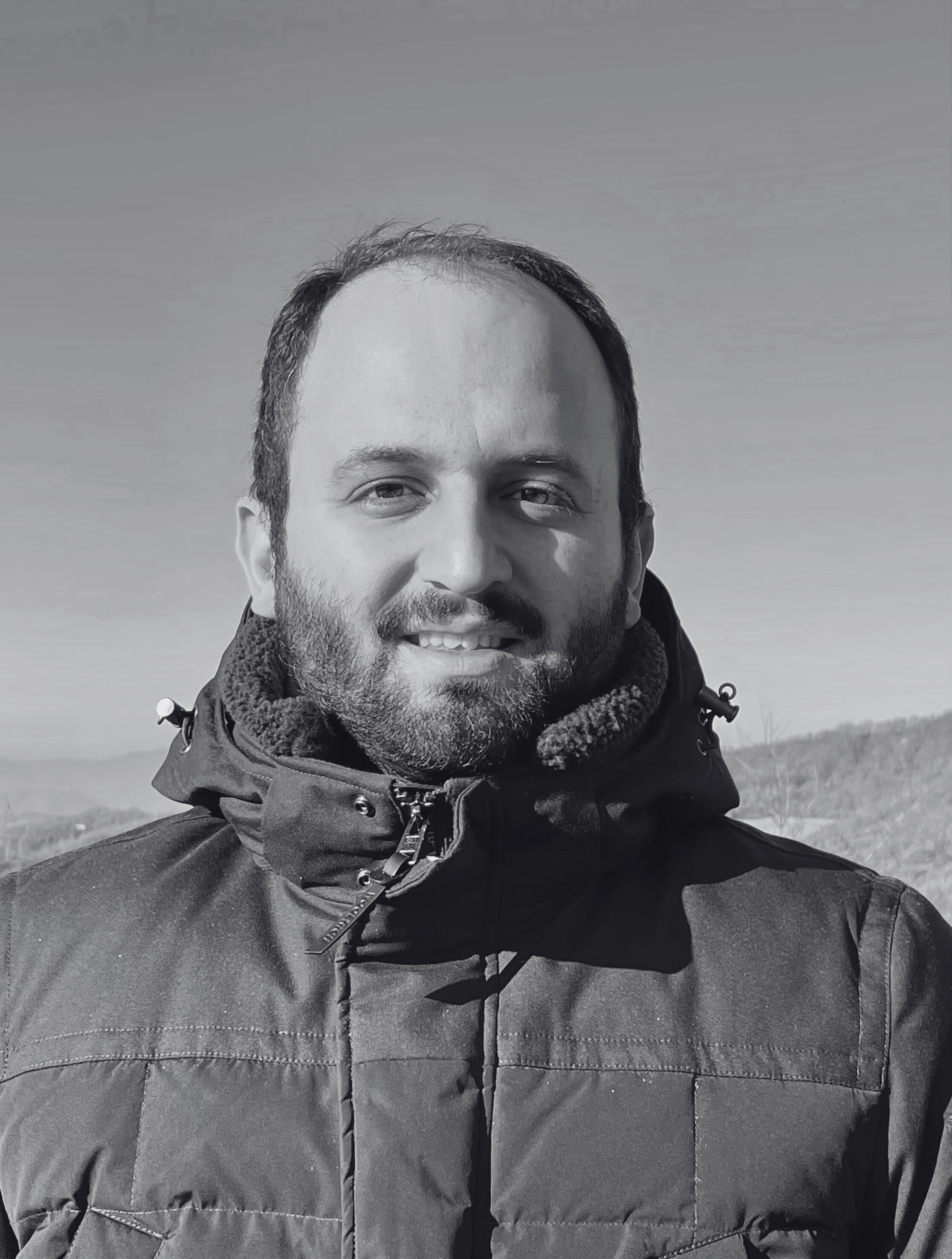}}]{Francesco Conti} 
received the Ph.D. degree in electronic engineering from the University of Bologna, Italy, in 2016. He is currently a Tenure-Track Assistant Professor with the DEI Department, University of Bologna. From 2016 to 2020, he held a research grant with the University of Bologna and a Post-Doctoral Researcher with ETH Zürich. His research is centered on hardware acceleration in ultra-low power and highly energy efficient platforms, with a particular focus on System-on-Chips for Artificial Intelligence applications. His research work has resulted in more than 70 publications in international conferences and journals and was awarded several times, including the 2020 IEEE TCAS-I Darlington Best Paper Award
\end{IEEEbiography}

\vspace{-10mm}

\begin{IEEEbiography}[{\includegraphics[width=1in,height=1.25in,clip,keepaspectratio]{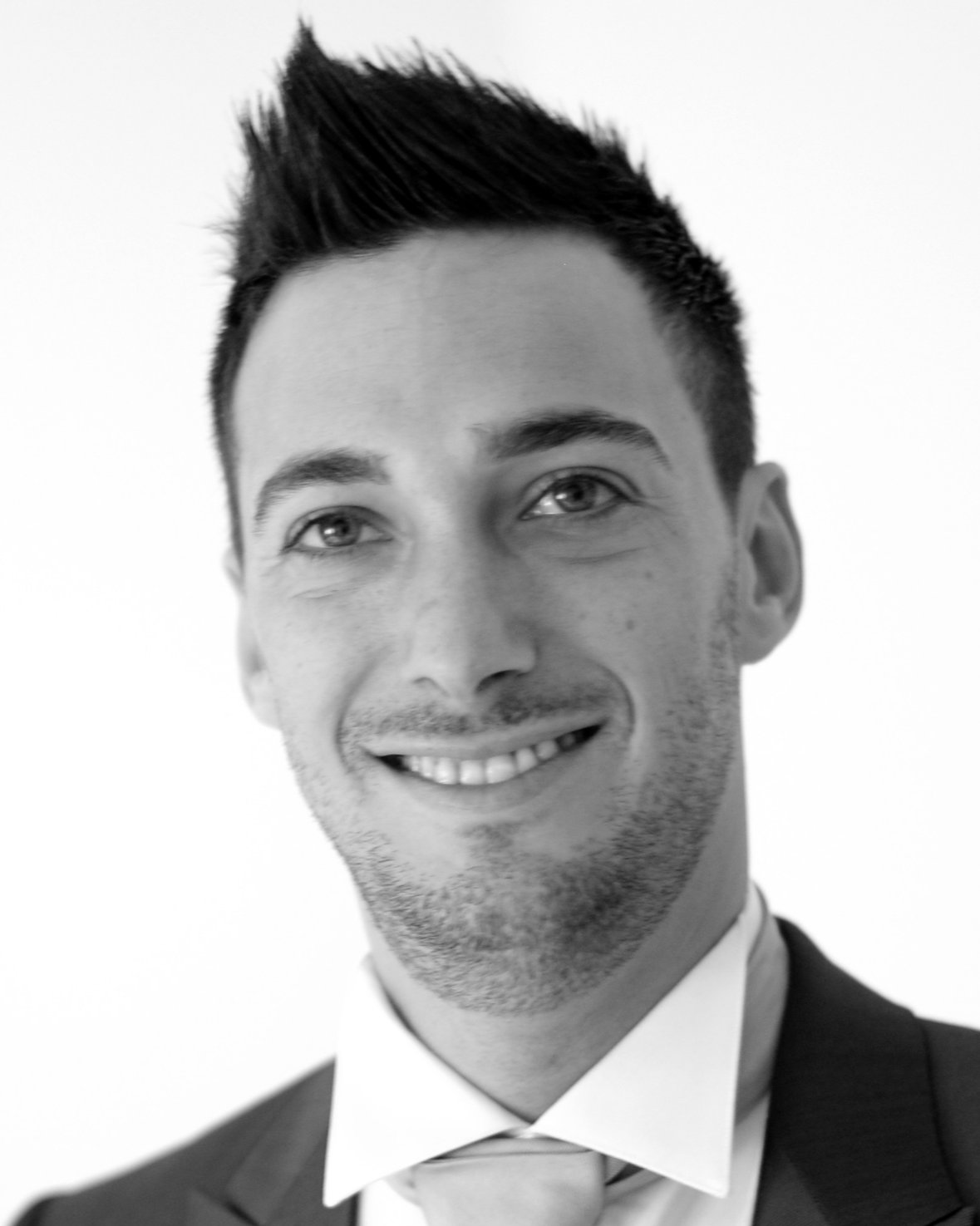}}]{Davide Rossi} received the Ph.D. degree from the University of Bologna, Bologna, Italy, in 2012. He has been a Post-Doctoral Researcher with the Department of Electrical, Electronic and Information Engineering “Guglielmo Marconi,” University of Bologna, since 2015, where he is currently an Associate Professor. His research interests focus on energy-efficient digital architectures. In this field, he has published more than 100 papers in international peer-reviewed conferences and journals. He is recipient of Donald O. Pederson Best Paper Award 2018, 2020 IEEE TCAS Darlington Best Paper Award, 2020 IEEE TVLSI Prize Paper Award.
\end{IEEEbiography}

%% file: Maestro_main.bbl
\begin{thebibliography}{10}
\providecommand{\url}[1]{#1}
\csname url@samestyle\endcsname
\providecommand{\newblock}{\relax}
\providecommand{\bibinfo}[2]{#2}
\providecommand{\BIBentrySTDinterwordspacing}{\spaceskip=0pt\relax}
\providecommand{\BIBentryALTinterwordstretchfactor}{4}
\providecommand{\BIBentryALTinterwordspacing}{\spaceskip=\fontdimen2\font plus
\BIBentryALTinterwordstretchfactor\fontdimen3\font minus \fontdimen4\font\relax}
\providecommand{\BIBforeignlanguage}[2]{{%
\expandafter\ifx\csname l@#1\endcsname\relax
\typeout{** WARNING: IEEEtran.bst: No hyphenation pattern has been}%
\typeout{** loaded for the language `#1'. Using the pattern for}%
\typeout{** the default language instead.}%
\else
\language=\csname l@#1\endcsname
\fi
#2}}
\providecommand{\BIBdecl}{\relax}
\BIBdecl

\bibitem{10529060}
J.~S. Letchumanan, S.~Gandhi \emph{et~al.}, ``A mechanically flexible 32-by-32-element pitch-matched ultrasound front-end transceiver with two-stage beamforming for 3d imaging,'' in \emph{2024 IEEE Custom Integrated Circuits Conference (CICC)}, 2024, pp. 1--2.

\bibitem{HU2024107401}
\BIBentryALTinterwordspacing
H.~Hu, C.~Hu \emph{et~al.}, ``Wearable ultrasound devices: An emerging era for biomedicine and clinical translation,'' \emph{Ultrasonics}, vol. 142, p. 107401, 2024. [Online]. Available: \url{https://www.sciencedirect.com/science/article/pii/S0041624X24001641}
\BIBentrySTDinterwordspacing

\bibitem{9185023}
X.~Yang, Y.~Zhou \emph{et~al.}, ``Wearable ultrasound-based decoding of simultaneous wrist/hand kinematics,'' \emph{IEEE Transactions on Industrial Electronics}, vol.~68, no.~9, pp. 8667--8675, 2021.

\bibitem{10208224}
J.~Zeng, Y.~Sheng \emph{et~al.}, ``Adaptive learning against muscle fatigue for a-mode ultrasound-based gesture recognition,'' \emph{IEEE Transactions on Instrumentation and Measurement}, vol.~72, pp. 1--10, 2023.

\bibitem{ManuelWearable}
M.~Eggimann, S.~Mach \emph{et~al.}, ``A risc-v based open hardware platform for always-on wearable smart sensing,'' in \emph{2019 IEEE 8th International Workshop on Advances in Sensors and Interfaces (IWASI)}, 2019, pp. 169--174.

\bibitem{BioWAP}
J.~Liu, Z.~Xie \emph{et~al.}, ``Biowap: A reconfigurable biomedical ai processor with adaptive processing for co-optimized accuracy and energy efficiency,'' in \emph{2024 IEEE Custom Integrated Circuits Conference (CICC)}, 2024, pp. 1--8.

\bibitem{VEGA}
D.~Rossi, F.~Conti \emph{et~al.}, ``4.4 a 1.3tops/w @ 32gops fully integrated 10-core soc for iot end-nodes with 1.7μw cognitive wake-up from mram-based state-retentive sleep mode,'' in \emph{2021 IEEE International Solid-State Circuits Conference (ISSCC)}, vol.~64, 2021, pp. 60--62.

\bibitem{8662605}
W.~Xia, Y.~Zhou \emph{et~al.}, ``Toward portable hybrid surface electromyography/a-mode ultrasound sensing for human–machine interface,'' \emph{IEEE Sensors Journal}, vol.~19, no.~13, pp. 5219--5228, 2019.

\bibitem{9872106}
Z.~Yin, H.~Chen \emph{et~al.}, ``A wearable ultrasound interface for prosthetic hand control,'' \emph{IEEE Journal of Biomedical and Health Informatics}, vol.~26, no.~11, pp. 5384--5393, 2022.

\bibitem{10557691}
A.~Bashatah, B.~Mukherjee \emph{et~al.}, ``Wearable ultrasound system using low-voltage time delay spectrometry for dynamic tissue imaging,'' \emph{IEEE Transactions on Biomedical Engineering}, vol.~71, no.~11, pp. 3232--3243, 2024.

\bibitem{UltrasoundPrediction}
J.~Mendez, R.~Murray \emph{et~al.}, ``A-mode ultrasound-based prediction of transfemoral amputee prosthesis walking kinematics via an artificial neural network,'' \emph{IEEE Transactions on Neural Systems and Rehabilitation Engineering}, vol.~PP, pp. 1--1, 02 2023.

\bibitem{MOHIT2024}
\BIBentryALTinterwordspacing
K.~Mohit, R.~Gupta \emph{et~al.}, ``A survey on the machine learning techniques for automated diagnosis from ultrasound images,'' \emph{Current Medical Imaging}, vol.~20, 2024. [Online]. Available: \url{https://www.sciencedirect.com/science/article/pii/S1573405624000055}
\BIBentrySTDinterwordspacing

\bibitem{7864441}
M.~Gautschi, P.~D. Schiavone \emph{et~al.}, ``Near-threshold risc-v core with dsp extensions for scalable iot endpoint devices,'' \emph{IEEE Transactions on Very Large Scale Integration (VLSI) Systems}, vol.~25, no.~10, pp. 2700--2713, 2017.

\bibitem{STM32L0}
\BIBentryALTinterwordspacing
S.~Microelectronics. (2019) Stm32l0 series. [Online]. Available: \url{https://www.st.com/en/microcontrollers-microprocessors/stm32l0-series.html}
\BIBentrySTDinterwordspacing

\bibitem{NDP250}
\BIBentryALTinterwordspacing
Syntiant. (2024) Syntiant. [Online]. Available: \url{https://www.syntiant.com/ndp250}
\BIBentrySTDinterwordspacing

\bibitem{STM32N6}
\BIBentryALTinterwordspacing
S.~Microelectronics. (2024) Stm32n6 series. [Online]. Available: \url{https://www.st.com/en/microcontrollers-microprocessors/stm32n6-series.html}
\BIBentrySTDinterwordspacing

\bibitem{AlifBalletto}
\BIBentryALTinterwordspacing
A.~Semiconductor. (2024) Balletto. [Online]. Available: \url{https://alifsemi.com/products/balletto/}
\BIBentrySTDinterwordspacing

\bibitem{Dabbelt2016VectorPF}
\BIBentryALTinterwordspacing
D.~P. Dabbelt, C.~Schmidt \emph{et~al.}, ``Vector processors for energy-efficient embedded systems,'' \emph{Proceedings of the Third ACM International Workshop on Many-core Embedded Systems}, 2016. [Online]. Available: \url{https://api.semanticscholar.org/CorpusID:16738646}
\BIBentrySTDinterwordspacing

\bibitem{Cavalcante2023SpatzCC}
\BIBentryALTinterwordspacing
M.~A. Cavalcante, M.~Perotti \emph{et~al.}, ``Spatz: Clustering compact risc-v-based vector units to maximize computing efficiency,'' \emph{ArXiv}, vol. abs/2309.10137, 2023. [Online]. Available: \url{https://api.semanticscholar.org/CorpusID:262054926}
\BIBentrySTDinterwordspacing

\bibitem{ARA}
M.~Cavalcante, F.~Schuiki \emph{et~al.}, ``Ara: A 1-ghz+ scalable and energy-efficient risc-v vector processor with multiprecision floating-point support in 22-nm fd-soi,'' \emph{IEEE Transactions on Very Large Scale Integration (VLSI) Systems}, vol.~28, no.~2, pp. 530--543, 2020.

\bibitem{YUN}
M.~Perotti, M.~Cavalcante \emph{et~al.}, ``Yun: An open-source, 64-bit risc-v-based vector processor with multi-precision integer and floating-point support in 65-nm cmos,'' \emph{IEEE Transactions on Circuits and Systems II: Express Briefs}, vol.~70, no.~10, pp. 3732--3736, 2023.

\bibitem{SPEED}
C.~Wang, C.~Fang \emph{et~al.}, ``Speed: A scalable risc-v vector processor enabling efficient multiprecision dnn inference,'' \emph{IEEE Transactions on Very Large Scale Integration (VLSI) Systems}, vol.~33, no.~1, pp. 207--220, 2025.

\bibitem{gemmini}
H.~Genc, S.~Kim \emph{et~al.}, ``Gemmini: Enabling systematic deep-learning architecture evaluation via full-stack integration,'' in \emph{2021 58th ACM/IEEE Design Automation Conference (DAC)}, 2021, pp. 769--774.

\bibitem{Tsunami}
S.~Kim, J.~Lee \emph{et~al.}, ``Tsunami: Triple sparsity-aware ultra energy-efficient neural network training accelerator with multi-modal iterative pruning,'' \emph{IEEE Transactions on Circuits and Systems I: Regular Papers}, vol.~69, no.~4, pp. 1494--1506, 2022.

\bibitem{Garofalo2023DARKSIDEAH}
\BIBentryALTinterwordspacing
A.~Garofalo, Y.~Tortorella \emph{et~al.}, ``Darkside: A heterogeneous risc-v compute cluster for extreme-edge on-chip dnn inference and training,'' \emph{IEEE Open Journal of the Solid-State Circuits Society}, vol.~2, pp. 231--243, 2023. [Online]. Available: \url{https://api.semanticscholar.org/CorpusID:252581131}
\BIBentrySTDinterwordspacing

\bibitem{Echoes}
M.~Sinigaglia, L.~Bertaccini \emph{et~al.}, ``Echoes: a 200 gops/w frequency domain soc with fft processor and i2s dsp for flexible data acquisition from microphone arrays,'' in \emph{2023 IEEE International Symposium on Circuits and Systems (ISCAS)}, 2023, pp. 1--5.

\bibitem{8240277}
A.~Wang, B.~Richards \emph{et~al.}, ``A 0.37mm2 lte/wi-fi compatible, memory-based, runtime-reconfigurable 2n3m5k fft accelerator integrated with a risc-v core in 16nm finfet,'' in \emph{2017 IEEE Asian Solid-State Circuits Conference (A-SSCC)}, 2017, pp. 305--308.

\bibitem{10509999}
H.~K. Sahu, E.~Sarkar \emph{et~al.}, ``Implementation of fft using low-precision floating point for rapid high precision mems readouts,'' in \emph{2023 IEEE Asia Pacific Conference on Circuits and Systems (APCCAS)}, 2023, pp. 173--177.

\bibitem{9926333}
L.~Tang, S.~Chen \emph{et~al.}, ``A high throughput hardware accelerator for fftw codelets: A first look,'' in \emph{2022 IEEE High Performance Extreme Computing Conference (HPEC)}, 2022, pp. 1--7.

\bibitem{STM32F7}
\BIBentryALTinterwordspacing
S.~Microelectronics. (2016) Stm32f7 series. [Online]. Available: \url{https://www.st.com/en/microcontrollers-microprocessors/stm32f7-series.html}
\BIBentrySTDinterwordspacing

\bibitem{TMS320F28379D}
\BIBentryALTinterwordspacing
T.~Instruments. (2025) Tms320f28379d. [Online]. Available: \url{https://www.ti.com/product/TMS320F28379D?keyMatch=TMS320F28379D&tisearch=universal_search&usecase=GPN}
\BIBentrySTDinterwordspacing

\bibitem{dsPIC33EP512MU814}
\BIBentryALTinterwordspacing
M.~Technology. (2022) dspic33ep512mu814. [Online]. Available: \url{https://www.microchip.com/en-us/product/dspic33ep512mu814}
\BIBentrySTDinterwordspacing

\bibitem{10.5555/1999263}
J.~L. Hennessy and D.~A. Patterson, \emph{Computer Architecture, Fifth Edition: A Quantitative Approach}, 5th~ed.\hskip 1em plus 0.5em minus 0.4em\relax San Francisco, CA, USA: Morgan Kaufmann Publishers Inc., 2011.

\bibitem{7924233}
N.~Stephens, S.~Biles \emph{et~al.}, ``The arm scalable vector extension,'' \emph{IEEE Micro}, vol.~37, no.~2, pp. 26--39, 2017.

\bibitem{RVV1}
\BIBentryALTinterwordspacing
R.-V.~C. 2022. (2022) Risc-v “v” vector extension, version 1.0. [Online]. Available: \url{https://github.com/riscvarchive/riscv-v-spec}
\BIBentrySTDinterwordspacing

\bibitem{Fugaku}
M.~Sato, ``The supercomputer “fugaku”,'' in \emph{2022 International Symposium on VLSI Design, Automation and Test (VLSI-DAT)}, 2022, pp. 1--1.

\bibitem{Vitruvius}
\BIBentryALTinterwordspacing
F.~Minervini, O.~Palomar \emph{et~al.}, ``Vitruvius+: An area-efficient risc-v decoupled vector coprocessor for high performance computing applications,'' \emph{ACM Trans. Archit. Code Optim.}, vol.~20, no.~2, Mar. 2023. [Online]. Available: \url{https://doi.org/10.1145/3575861}
\BIBentrySTDinterwordspacing

\bibitem{9567768}
A.~Gonzalez, J.~Zhao \emph{et~al.}, ``A 16mm2 106.1 gops/w heterogeneous risc-v multi-core multi-accelerator soc in low-power 22nm finfet,'' in \emph{ESSCIRC 2021 - IEEE 47th European Solid State Circuits Conference (ESSCIRC)}, 2021, pp. 259--262.

\bibitem{eightCore}
C.~Schmidt, J.~Wright \emph{et~al.}, ``4.3 an eight-core 1.44ghz risc-v vector machine in 16nm finfet,'' in \emph{2021 IEEE International Solid-State Circuits Conference (ISSCC)}, vol.~64, 2021, pp. 58--60.

\bibitem{10631150}
T.~Zhao and Z.~Ye, ``Zerovex: A scalable and high-performance risc-v vector processor core for embedded systems,'' in \emph{2024 IEEE 35th International Conference on Application-specific Systems, Architectures and Processors (ASAP)}, 2024, pp. 32--33.

\bibitem{CVA6}
F.~Zaruba and L.~Benini, ``The cost of application-class processing: Energy and performance analysis of a linux-ready 1.7-ghz 64-bit risc-v core in 22-nm fdsoi technology,'' \emph{IEEE Transactions on Very Large Scale Integration (VLSI) Systems}, vol.~27, no.~11, pp. 2629--2640, 2019.

\bibitem{TORTORELLA2023122}
\BIBentryALTinterwordspacing
Y.~Tortorella, L.~Bertaccini \emph{et~al.}, ``Redmule: A mixed-precision matrix–matrix operation engine for flexible and energy-efficient on-chip linear algebra and tinyml training acceleration,'' \emph{Future Generation Computer Systems}, vol. 149, pp. 122--135, 2023. [Online]. Available: \url{https://www.sciencedirect.com/science/article/pii/S0167739X23002546}
\BIBentrySTDinterwordspacing

\bibitem{5669293}
E.~E. Swartzlander and H.~H. Saleh, ``Fft implementation with fused floating-point operations,'' \emph{IEEE Transactions on Computers}, vol.~61, no.~2, pp. 284--288, 2012.

\bibitem{7123672}
A.~Kaivani and S.~Ko, ``Floating-point butterfly architecture based on binary signed-digit representation,'' \emph{IEEE Transactions on Very Large Scale Integration (VLSI) Systems}, vol.~24, no.~3, pp. 1208--1211, 2016.

\bibitem{Pedram2014AHE}
\BIBentryALTinterwordspacing
A.~Pedram, J.~D. McCalpin \emph{et~al.}, ``A highly efficient multicore floating-point fft architecture based on hybrid linear algebra/fft cores,'' \emph{Journal of Signal Processing Systems}, vol.~77, pp. 169 -- 190, 2014. [Online]. Available: \url{https://api.semanticscholar.org/CorpusID:9264564}
\BIBentrySTDinterwordspacing

\bibitem{biowulpus}
\BIBentryALTinterwordspacing
S.~Frey, V.~J. Kartsch \emph{et~al.}, ``A wearable ultra-low-power semg-triggered ultrasound system for long-term muscle activity monitoring,'' \emph{2023 IEEE International Ultrasonics Symposium (IUS)}, pp. 1--4, 2023. [Online]. Available: \url{https://api.semanticscholar.org/CorpusID:261706192}
\BIBentrySTDinterwordspacing

\bibitem{biogap}
S.~Frey, M.~Guermandi \emph{et~al.}, ``Biogap: a 10-core fp-capable ultra-low power iot processor, with medical-grade afe and ble connectivity for wearable biosignal processing,'' 07 2023, pp. 1--7.

\bibitem{wulpus}
S.~Frey, S.~Vostrikov \emph{et~al.}, ``Wulpus: a wearable ultra low-power ultrasound probe for multi-day monitoring of carotid artery and muscle activity,'' in \emph{2022 IEEE International Ultrasonics Symposium (IUS)}, 2022, pp. 1--4.

\bibitem{nRF52811}
\BIBentryALTinterwordspacing
N.~Semiconductor. (2019) nrf52811. [Online]. Available: \url{https://www.nordicsemi.com/Products/nRF52811}
\BIBentrySTDinterwordspacing

\bibitem{MSP430}
\BIBentryALTinterwordspacing
T.~Instruments. (2016) Msp430. [Online]. Available: \url{https://www.ti.com/microcontrollers-mcus-processors/msp430-microcontrollers/overview.html}
\BIBentrySTDinterwordspacing

\bibitem{nRF52832}
\BIBentryALTinterwordspacing
N.~Semiconductor. (2016) nrf52832. [Online]. Available: \url{https://www.nordicsemi.com/Products/nRF52832}
\BIBentrySTDinterwordspacing

\bibitem{8106976}
P.~Davide~Schiavone, F.~Conti \emph{et~al.}, ``Slow and steady wins the race? a comparison of ultra-low-power risc-v cores for internet-of-things applications,'' in \emph{2017 27th International Symposium on Power and Timing Modeling, Optimization and Simulation (PATMOS)}, 2017, pp. 1--8.

\bibitem{Snitch}
F.~Zaruba, F.~Schuiki \emph{et~al.}, ``Snitch: A tiny pseudo dual-issue processor for area and energy efficient execution of floating-point intensive workloads,'' \emph{IEEE Transactions on Computers}, vol.~70, no.~11, pp. 1845--1860, 2021.

\bibitem{10.1109/TC.2023.3329930}
\BIBentryALTinterwordspacing
T.~Benz, M.~Rogenmoser \emph{et~al.}, ``A high-performance, energy-efficient modular dma engine architecture,'' \emph{IEEE Trans. Comput.}, vol.~73, no.~1, p. 263–277, Nov. 2023. [Online]. Available: \url{https://doi.org/10.1109/TC.2023.3329930}
\BIBentrySTDinterwordspacing

\bibitem{Cooley1965AnAF}
\BIBentryALTinterwordspacing
J.~W. Cooley and J.~W. Tukey, ``An algorithm for the machine calculation of complex fourier series,'' \emph{Mathematics of Computation}, vol.~19, pp. 297--301, 1965. [Online]. Available: \url{https://api.semanticscholar.org/CorpusID:121744946}
\BIBentrySTDinterwordspacing

\bibitem{bertaccini2021buffer}
L.~Bertaccini, L.~Benini \emph{et~al.}, ``To buffer, or not to buffer? a case study on fft accelerators for ultra-low-power multicore clusters,'' in \emph{2021 IEEE 32nd International Conference on Application-specific Systems, Architectures and Processors (ASAP)}.\hskip 1em plus 0.5em minus 0.4em\relax IEEE, 2021, pp. 1--8.

\bibitem{8766229}
``Ieee standard for floating-point arithmetic,'' \emph{IEEE Std 754-2019 (Revision of IEEE 754-2008)}, pp. 1--84, 2019.

\end{thebibliography}
